\documentclass[11pt]{article}

\usepackage{amssymb,amsmath}
\usepackage{latexsym}
\usepackage{mathrsfs}
\usepackage[%
             colorlinks=true,urlcolor=blue,linkcolor=blue,
             pdfpagelabels=true,hypertexnames=true,
            plainpages=false,naturalnames=false,
             ]{hyperref}
\usepackage{verbatim}
\usepackage{bm}
\usepackage{cite}

\def\bea{\begin{eqnarray}}
\def\eea{\end{eqnarray}}
\def\be{\begin{equation}}
\def\ee{\end{equation}}
\def\ba{\begin{array}}
\def\ea{\end{array}}
\def\lsim{\hbox{ \raise.35ex\rlap{$<$}\lower.6ex\hbox{$\sim$}\ }}
\def\gsim{\hbox{ \raise.35ex\rlap{$>$}\lower.6ex\hbox{$\sim$}\ }}

\def\nn{\nonumber}

\begin{document}

\setlength\arraycolsep{2pt}

\renewcommand{\theequation}{\arabic{section}.\arabic{equation}}
\setcounter{page}{1}

\begin{titlepage}

\rightline{\footnotesize{KCL-PH-TH/2012-43}} \vspace{-0.2cm}

\begin{center}

\vskip 1.0 cm

{\LARGE  \bf  Ef\mbox{}fective f\mbox{}ield theory of weakly coupled inf\mbox{}lationary models }

\vskip 1.0cm

{\large
Rhiannon Gwyn$^{a}$, Gonzalo A. Palma$^{b}$, Mairi Sakellariadou$^{c}$ \\ and Spyros Sypsas$^{c}$
}

\vskip 0.5cm

{\it
$^{a}$Max-Planck-Institut f\"ur Gravitationsphysik, Albert-Einstein-Institut\\
\mbox{Am M\"uhlenberg 1, D-14476 Potsdam, Germany}
\\
$^{b}$Physics Department, FCFM, Universidad de Chile \\ \mbox{Blanco Encalada 2008, Santiago, Chile}
\\
$^{c}$Department of Physics, King's College London \\ \mbox{Strand, London WC2R 2LS, U.K.}
}

\vskip 1.5cm

\end{center}

\begin{abstract}

The application of Effective Field Theory (EFT) methods to inflation has taken a central role in our current understanding of the very early universe. The EFT perspective has been particularly useful in analyzing the self-interactions determining the evolution of co-moving curvature perturbations (Goldstone boson modes) and their influence on low-energy observables. However, the standard EFT formalism, to lowest order in spacetime differential operators, does not provide the most general parametrization of a theory that remains weakly coupled throughout the entire low-energy regime. Here we study the EFT formulation by including spacetime differential operators implying a scale dependence of the Goldstone boson self-interactions and its dispersion relation. These operators are shown to arise naturally from the low-energy interaction of the Goldstone boson with heavy fields that have been integrated out. We find that the EFT then stays weakly coupled all the way up to the cutoff scale at which ultraviolet degrees of freedom become operative.  This opens up a regime of new physics where the dispersion relation is dominated by a quadratic dependence on the momentum $\omega \sim p^2$. In addition, provided that modes crossed the Hubble scale within this energy range, the predictions of inflationary observables ---including non-Gaussian signatures--- are significantly affected by the new scales characterizing it. 

\end{abstract}

\end{titlepage}

\newpage

\hypersetup{linktocpage}
\tableofcontents

\section{Introduction \& summary}\label{intro}
\setcounter{equation}{0}

Effective field theory (EFT) constitutes a powerful and unified scheme to study the possible effects of unknown ultraviolet (UV) degrees of freedom on the low-energy evolution of curvature perturbations during inflation~\cite{Cheung:2007st, Weinberg:2008hq}. By employing general symmetry arguments, it is possible to deduce the most general EFT parametrizing the low-energy dynamics of curvature perturbations on a Friedmann-Lema\^itre-Robertson-Walker (FLRW) background. The benefit of adopting such a  perspective is manyfold: 
Firstly, it allows one to focus the study of inflation on the dynamics of curvature fluctuations ---which are ultimately responsible for any low-energy observable today--- relegating the model-dependent background dynamics to a subsidiary plane. 
Secondly, it offers a simple and intuitive interpretation of curvature perturbations as Goldstone boson modes, emerging as a consequence of the spontaneous breaking of time translation invariance. This, in turn, allows the application of well known techniques, both perturbative and non-perturbative, to analyze various cosmological observables, such as $n$-point correlation functions. 
Thirdly, it offers an explicit parametrization of all the relevant couplings in the low-energy evolution of curvature perturbations, simplifying the study of the relations between field-operators appearing at different orders in perturbation theory. All of these characteristics have invigorated the study of inflation, allowing for a model-independent approach to the analysis of the infrared (IR) inflationary observables accessible today~\cite{Cheung:2007sv, Senatore:2009gt, Senatore:2009cf, Senatore:2010jy, Senatore:2010wk,Khosravi:2012qg, Creminelli:2010qf, Baumann:2011dt, Baumann:2011su, Baumann:2011nk,Baumann:2011nm,  LopezNacir:2011kk, Baumann:2011ws, Achucarro:2012sm, Senatore:2012wy}.

In the particular case where the inflationary perturbations are generated by a single scalar degree of freedom, a fully satisfactory effective field theory of their dynamics was first derived by Cheung {\it et al.} in ref.~\cite{Cheung:2007st}. The basic procedure adopted there was simple and intuitive: First, one postulates as a background an FLRW geometry that breaks time translation invariance but leaves the spatial diffeomorphisms untouched. Then, the spontaneous breaking of time translation invariance necessarily results in an extra scalar degree of freedom which reveals itself as a longitudinal polarization of the metric. In unitary gauge, this scalar degree of freedom may be identified with the $00$-component of the metric $\delta g^{00} \equiv 1 + g^{00}  $ and, to lowest order in spacetime differential operators, its effective action takes the form
\bea
S_{\rm EFT} \! &=& \!\! \int \!\! d^3xdt \sqrt{-g} \bigg[ \frac{M_{\rm Pl}^2}{2} R + M_{\rm Pl}^2 \dot H g^{00} - M_{\rm Pl}^2 (3 H^2 + \dot H) \nn \\
&&  + \frac{1}{2!} M_2^4(t) ( g^{00} + 1)^2  + \frac{1}{3!} M_3^4(t) (g^{00} + 1)^3 + \cdots    \nn \\ 
&& - \frac{\bar M_1^3(t)}{2} (g^{00} + 1) \delta K^{\mu}{}_{\mu} - \frac{\bar M_2^2(t)}{2}  \delta K^{\mu}{}_{\mu}{}^2 - \frac{\bar M_3^2(t)}{2} \delta K^{\mu}{}_{\nu}   \delta K^{\nu}{}_{\mu}   + \cdots \bigg]
,  \label{starting-action}
\eea
where $M_{\rm Pl}$ stands for the Planck mass, $H = \dot a / a$ is the expansion rate defined in terms of the scale factor $a(t)$, and $\delta K^{\mu}{}_{\nu}$ denotes the perturbed extrinsic curvature of spatial foliations at a fixed time. In the previous expression, the second and third terms are required so that the background Friedmann equations are satisfied, independently of the fluid driving inflation. The $M_n(t)$ and $\bar M_n(t)$ coefficients are functions of time only, and parametrize the effects on the low-energy dynamics due to the unknown UV physics. They should therefore be regarded as undetermined parameters of the theory.  For instance, setting $M_n(t) = \bar M_n(t) = 0$ corresponds to the standard case derived from an action for a single canonical scalar field theory with a flat potential~\cite{Linde:1981mu, Albrecht:1982wi, Mukhanov:1981xt}. It should be clear that the action (\ref{starting-action}) does not reveal the energy scale $\Lambda_{\rm UV}$ at which ultraviolet degrees of freedom start playing a relevant role in the inflationary dynamics. Nevertheless, one typically expects the coefficients $M_n(t)$ and $\bar M_n(t)$ to depend on a combination of scales involving $\Lambda_{\rm UV}$, $M_{\rm Pl}$ and the background quantities $H$ and $\dot H$.

The Goldstone boson field $\pi  (x)$ may be introduced as the adiabatic field fluctuation along the time direction, in such a way that the spontaneously broken time diffeomorphism $t \to t + \xi^0$ is non-linearly realized through the complementary field transformation $\pi \to \pi - \xi^0$ (see Section \ref{Goldstone-boson-computations} for more details). At lowest order in perturbation theory, the Goldstone boson is simply related to the co-moving curvature perturbation $\zeta (x)$ by  $\pi (x) = - \zeta (x) / H$. 
In terms of the Goldstone boson, the effects due to $M_2^4$ are already relevant at the free field theory level, where its appearance results in a reduction of  the speed of sound $c_{\rm s}$ at which fluctuations propagate. Concretely, one finds that the propagation of $\pi$-modes is characterized by a dispersion relation given by
$$
\omega(p) = c_{\rm s} p,  \qquad \textrm{with} \qquad \frac{1}{c_{\rm s}^2} = 1 + \frac{2 M_2^4 }{ M_{\rm Pl}^2 | \dot H|} , 
$$
where $p$ is the momentum carried by the Goldstone boson quanta. Crucially, because $1 + g^{00}$ depends non-linearly on $\pi$, a suppression of the speed of sound inevitably implies the existence of higher-order interactions with strengths proportional to $M_2^4 = (c_{\rm s}^{-2} - 1)M_{\rm Pl}^2 | \dot H|/2$. 
As a consequence, a suppression of the speed of sound at the free field theory level also implies the appearance of non-trivial cubic interactions leading to potentially large levels of equilateral non-Gaussianity characterized by an $f_{\rm NL}^{\rm (eq)}$ parameter of the form~\cite{Cheung:2007st, Senatore:2009gt}:
\be
f_{\rm NL}^{\rm (eq)} \sim  \frac{1}{4 c_{\rm s}^2} . \label{non-Gaussianity-c-s}
\ee
Given that $M_2$ is a mass scale related to the unknown UV physics, unsuppressed by the symmetries of the background, observation of large non-Gaussianity is quite possible in future cosmological probes. Current observational bounds on $f_{\rm NL}^{\rm (eq)} $ imply an approximate lower bound on $c_{\rm s}$ given by $c_{\rm s} > 0.01$~\cite{Senatore:2009gt, Komatsu:2010fb}.
However, the non-linear interactions in action (\ref{starting-action}) induced by $c_{\rm s} < 1$ are non-unitary and, as such, they imply that the theory becomes strongly coupled at an energy scale sensitive to $c_{\rm s}$, given by\footnote{See Appendix E of \cite{Baumann:2011su} for a detailed derivation. Note that the expression given in \cite{Baumann:2011su} differs by a factor of $4 \pi$ from that given in \cite{Cheung:2007st}.} 
$$ \Lambda_{\rm s.c.}^4 = 4 \pi M_{\rm Pl}^2 |\dot H | c_{\rm s}^5 (1 - c_{\rm s}^2)^{-1}. $$  
Since a reduction in the speed of sound decreases the value of this strong coupling scale, there is a lower bound on how small $c_{\rm s}$ can be or, equivalently, on how large non-Gaussian signatures are allowed to be without rendering the effective description invalid. A simple estimation, taking into account the observed value of the amplitude of the primordial power spectrum $\mathcal P_\zeta$ of curvature perturbations, implies the constraint $c_{\rm s} > 0.01$ in order to avoid strong coupling of the EFT at energy scales relevant at Hubble horizon\footnote{Although horizon is a widely used term in expressions like ''horizon crossing", throughout the paper we will use ``Hubble horizon" or ``Hubble crossing" instead, to avoid confusion with the particle horizon. See e.g. p.40 of \cite{Mukhanov:2005sc} for a discussion.} crossing~\cite{Cheung:2007st}. We thus see that this result is consistent with the previous observational constraint.

However, the previous result implies that for small values of the speed of sound ($c_{\rm s}^2 \ll 1$), Goldstone boson modes described by (\ref{starting-action}) may appear strongly coupled at energies well below the cutoff energy scale $\Lambda_{\rm UV}$ at which UV degrees of freedom become excited. This fact reflects a limitation of the EFT (\ref{starting-action}) to consistently parametrize the class of theories that remain weakly coupled as the energy increases up to $\Lambda_{\rm UV}$. In ref.~\cite{Baumann:2011su}, Baumann and Green addressed this issue by studying weakly coupled completions of (\ref{starting-action}). They pointed out that any field theoretical description remaining weakly coupled all the way up to the symmetry breaking scale will involve either new degrees of freedom or an energy regime with new physics, in such a way that non-unitary operators stay suppressed. This {\it new physics} regime was found to be characterized by a modified dispersion relation with quadratic momentum dependence
$$
\omega (p) \propto p^2 , 
$$ 
which can equivalently be seen to lift the strong coupling scale to a value larger than $\Lambda_{\rm UV}$, so that one obtains a low-energy effective description of the system in terms of a weakly coupled Goldstone boson.

The analysis of~\cite{Baumann:2011su} did not address the more general problem of constructing an EFT fully consistent with the new physics regime, in such a way that the modified dispersion relation is non-linearly realized to all orders in perturbation theory.
The aim of the present work is to take a step forward in this direction. Our main emphasis is that the new physics regime is fully incorporated within the EFT formalism of~\cite{Cheung:2007st}, and a UV-completion is not needed to keep the theory weakly coupled. This is achieved by allowing action (\ref{starting-action}) to incorporate extra scale-dependent operators acting on the fields. Crucially, we find that the action for the Goldstone boson requires a non-trivial modification of its higher-order interactions, implying novel effects on the prediction of inflationary observables. Concretely, we show that, in its minimal version, the new physics regime requires a generalization of the EFT action (\ref{starting-action}) of the form
\bea
S_{\rm EFT} \!\! &=& \!\! \int \!\! d^3xdt \sqrt{-g} \bigg\{  \frac{M_{\rm Pl}^2}{2} R + M_{\rm Pl}^2 \dot H g^{00} - M_{\rm Pl}^2 (3 H^2 + \dot H)  +  \frac{M_2^4 }{2!} (1+ g^{00})   \frac{M^2 }{ M^2 -  \nabla^2/a^2} (1+g^{00})  \nn \\
&&
+  \frac{M_3^4 }{3!}  (1+ g^{00})   \frac{M^2 }{ M^2 - \nabla^2/a^2} \left [ (1+ g^{00})   \frac{M^2 }{ M^2 -  \nabla^2/a^2}  (1+g^{00})\right ]  + \cdots  \bigg\} ,  \label{modified-action-2}
\eea
where $\nabla^2 \equiv \delta^{ i j } \partial_i \partial_j$ is the Laplacian operator and $M$ is a mass scale parametrizing the new physics regime. This {\it minimal version} is found to be consistent with a  single field theory coupled to a set of heavy fields that are only excited at very high energies (in fact, at an energy that may be much larger than $M$). 
This may readily be understood by interpreting the operators $ (M^2 - \nabla^2/a^2)^{-1}$ as propagators characterizing states of mass $M$ interacting with the inflaton. Note that this is quite a natural interpretation of the scale-dependent interactions. All UV completions of gravity predict a vast variety of scalar fields (e.g. moduli) that are constrained to be heavy in order to be compatible with low-energy phenomenology. Since one expects that the UV physics is responsible for inflation, logically one also anticipates that the inflaton would be coupled to these heavy fields. As we shall see, the new physics regime appears whenever the Laplacian $\nabla^2$ dominates over the mass $M$ in 
the operator $(M^2 - \nabla^2/a^2)^{-1}$, forcing us to consider the formal ---but consistent--- expansion:
\be
\frac{1}{M^2 - \tilde\nabla^2 } = - \frac{1}{\tilde\nabla^2}  - \frac{M^2}{\tilde\nabla^4} - \cdots , \label{new-physics-expansion}
\ee
where $\tilde\nabla \equiv \nabla/a$.
We shall justify this expansion in Section~\ref{EFT-new-physics}, where we show that it emerges as a consequence of the non-relativistic nature of the theory, 
i.e. the breaking of time translation invariance, which is at the heart of the EFT formulation of inflation. As we shall demonstrate, these modifications imply that the EFT remains weakly coupled all the way up to the energy scale where the integrated UV degrees of freedom become operative, in agreement with the analysis of ref.~\cite{Baumann:2011su}. We will argue that these effects are conceivably generic in the sense that they consistently capture the gradual way in which UV corrections start to take over until the theory may need to be explicitly completed.

Further, we show that the generalized effective action (\ref{modified-action-2}) implies a non-trivial dependence of the inflationary observables on the parameters of the theory. To be more precise, if the modes relevant  for current observables crossed the Hubble scale within the new physics regime (that is, if for these modes $\omega \simeq H$ when $\omega \propto p^2$) the power spectrum $\mathcal P_{\zeta} $, the tensor to scalar ratio $r$ and the $f_{\rm NL}$ parameter are found to be given by
$$
\mathcal P_{\zeta}  \sim \frac{2.7}{100} \frac{H^2}{M_{\rm Pl}^2 \epsilon} \sqrt{\frac{M}{H c_{\rm s}}} , \qquad   r \sim 7.6 \epsilon  \sqrt{\frac{H c_{\rm s} }{M} } ,  \qquad f_{\rm NL} \sim \frac{M }{H c_{\rm s}} ,  \label{predictions-0}
$$
where $\epsilon=- \dot H / H^2$ is the usual slow-roll parameter. These expressions  differ from those derived from the standard effective field theory (\ref{starting-action}) and constitute a new parametrization of the inflationary observables (such as the energy scale of inflation) distinct from others found in the recent literature, therefore inviting us to consider more carefully the way in which future data should be interpreted.

Although the EFT formalism developed in ref.~\cite{Cheung:2007st} is entirely general, the bulk of the existing literature has focussed on analyzing inflation through the specific action (\ref{starting-action}), which corresponds to the long wavelength limit of (\ref{modified-action-2}). In this sense, our work emphasizes the strength of the formalism developed in ref.~\cite{Cheung:2007st} to study inflation more generally, by identifying new higher derivative operators that can capture effects of heavy fields coupled to the inflaton, consistent with the EFT formalism ~\cite{Cheung:2007st}. However, we stress that other parametric regimes of the EFT formalism, beyond that offered by (\ref{starting-action}), have already been studied in the past (see for instance refs.~\cite{Cheung:2007st, Senatore:2009gt, Senatore:2010jy}).

We have organized this article in the following way: In Section~\ref{Goldstone-boson-computations} we begin by summarizing the EFT formalism of ref.~\cite{Cheung:2007st}, focusing on the relation between the Goldstone boson and co-moving curvature perturbations. In Section~\ref{EFT-new-physics} we present and discuss the basic extension of the effective action to parametrize the new physics regime, including a discussion of the strong coupling scale.
Then, in Section~\ref{integration-1-massive-field}, we show how this form of the action emerges from the simple case in which a single heavy degree of freedom is integrated out (a more general treatment, where several heavy degrees of freedom are integrated out, is presented in Appendix~\ref{several-massive-fields}). There we also discuss the validity and consistency of the expansion (\ref{new-physics-expansion}) to parametrize the effects of UV-degrees of freedom on the low-energy dynamics of curvature perturbations.
In Section~\ref{Implications-inflation} we discuss the implications of the extended EFT for inflationary observables. We show that, indeed, the scales parametrizing the new physics regime modify the computation of two and three-point correlation functions, implying a re-interpretation of the parametrization of non-Gaussianity in terms of the speed of sound (\ref{non-Gaussianity-c-s}), or other Lagrangian operators \cite{Senatore:2009gt}, which is given by the standard EFT perspective.
Finally, in Section~\ref{section-conclusions} we present our concluding remarks.


\section{Review of the standard EFT formalism} \label{Goldstone-boson-computations}
\setcounter{equation}{0}

Here we review the EFT formalism developed in \cite{Cheung:2007st}, written in terms of the Goldstone boson $\pi$ and its relation, via gauge transformations, to the co-moving curvature perturbation $\zeta$. Readers familiar with this material may skip this discussion and go directly to Section \ref{EFT-new-physics}. 

Our starting point is to consider a quasi de Sitter time-dependent background that breaks time diffeomorphisms but keeps spatial ones. To this end, it is useful to adopt the Arnowitt-Deser-Misner (ADM) decomposition of the four-dimensional spacetime\cite{Arnowitt:1962hi} through the metric
\begin{equation}
ds^2=-N^2dt^2+\gamma_{ij}(N^idt+dx^i)(N^jdt+dx^j) , \label{ADM-metric}
\end{equation}
where $N$ and $N^i$ are the lapse and shift functions ---to be treated as Lagrange multipliers--- and $\gamma_{ij}$ is the induced metric on the 3-D spatial foliations. In terms of these quantities, the components of the metric $g_{\mu \nu}$ and its inverse $g^{\mu \nu}$ are given by
\begin{equation}
\begin{split}
g_{00}&=-N^2+\gamma_{ij}N^iN^j,\quad g_{0i}=\gamma_{ij}N^j,\quad g_{ij} = \gamma_{ij}, \\
g^{00}&=-\dfrac{1}{N^2},\quad g^{0i}=\dfrac{N^i}{N^2}\quad g^{ij}=\gamma^{ij}-\dfrac{N^iN^j}{N^2},
\end{split} \label{ADM-parametrization}
\end{equation}
where $\gamma^{ij}$ is the inverse of $\gamma_{ij}$.
Throughout this article we focus our attention on the case where primordial perturbations are due to a single scalar degree of freedom. In order to compute observables such as correlation functions, it is convenient to choose a gauge. There are two gauges that are particularly useful for this goal: One is the co-moving or unitary gauge, where the scalar degree of freedom corresponds to the co-moving curvature perturbation $\zeta (x, t)$ parametrizing inhomogeneous fluctuations of the flat spatial metric as:
\begin{equation} \label{zgauge}
g_{ij}=a^2(t)e^{2\zeta}\delta_{ij} \ .
\end{equation}
If we were considering a model of inflation with a scalar field $\phi$ playing the role of the inflaton, in this gauge we would have that its fluctuations are set to zero $\delta \phi = 0$. The other gauge we consider is the flat gauge, where the spatial metric corresponds to pure background:
\begin{equation} \label{pgauge}
g_{ij}=a^2(t)\delta_{ij} \ .
\end{equation}
In this case, the scalar degree of freedom appears explicitly in the matter sector of the theory, which in the present case does not need to be specified. The transformation between the two gauges is a spacetime re-parametrization \cite{Maldacena:2002vr}. Indeed, in order to go from \eqref{pgauge} to \eqref{zgauge} one performs a time re-parametrization of the form:
\be \label{timeshift}
t\rightarrow \tilde{t}=t+\pi(t,x).
\ee 
This introduces the Goldstone boson field $\pi (t,x)$ as a mean of parametrizing scalar perturbations in the flat gauge (\ref{pgauge}).
The scale factor $a(t)$ changes as $a(t)= a(\tilde{t}-\pi(t,x))$ which results in a Weyl rescaling $a(t) \rightarrow a (t) e^{\zeta(t,x)}$.  Thus, to first order in $\pi$, one has $a (t) = a(\tilde t - \pi) =  a(\tilde t ) e^{- H \pi}$, from which one can deduce the relation $\zeta = - H \pi$.  For the computation to second and higher order one has to iterate the Taylor expansion and compute $a(t)=a[\tilde{t}-\pi(\tilde{t}-\pi(t,x),x)]$ and so on. For example the result to second order is 
\be 
\begin{split}
a[\tilde t-\pi(\tilde t-\pi(t,x),x)] & =a(\tilde t)-  \dot a(\tilde t) \pi(\tilde t-\pi(t,x),x)+\frac{1}{2}\ddot a (\tilde t) [\pi(\tilde t-\pi(t,x),x)]^2 \ ,
\end{split}
\ee
which gives~\cite{Cheung:2007sv}
\be \label{zofpi}
\zeta^{(2)}(\pi)=-H \pi + H \pi \dot\pi + \frac{1}{2}\dot H \pi^2,
\ee
where we used $\pi(\tilde t-\pi)=\pi(\tilde t)+\mathcal{O}(\pi^2)$ and the definition of the Hubble parameter $H$. Because of this time re-parametrization the metric picks up non-diagonal terms starting from second order in $\pi$. These extra terms can be eliminated by performing a space re-parametrization with a parameter that is a series in $\pi$ starting at second order, an operation which results in more complicated expressions for $\zeta(\pi)$, involving spatial derivatives \cite{Maldacena:2002vr}. The variable $\zeta$ in the comoving gauge stays constant outside the Hubble horizon;  its correlators are the gauge invariant observables one wishes to compute. 
Alternatively, as we will see below, the flat gauge can be very useful from a computational point of view, since there exist physical limits where the problem simplifies considerably and one is able to draw powerful qualitative and quantitative conclusions.

As stated in the Introduction, the effective field theory describing a single scalar degree of freedom in the unitary gauge, where there are only metric fluctuations, takes the form
\bea
S_{\rm EFT} &=& \int d^3xdt \sqrt{-g} \bigg[ \frac{M_{\rm Pl}^2}{2} R + M_{\rm Pl}^2 \dot H g^{00} - M_{\rm Pl}^2 (3 H^2 + \dot H) \nn \\
&&  + \frac{1}{2!} M_2^4(t) (1+ g^{00})^2  + \frac{1}{3!} M_3^4(t) (1+ g^{00})^3 + \cdots  \bigg]   , \label{eft-action-scalar-sector}
\eea
where we have ignored contributions coming from perturbations of the extrinsic curvature $\delta K^{\mu}{}_{\nu}$.
 In order to go to the flat gauge one can invert the aforementioned time re-parametrization \eqref{timeshift} which may be thought of as the St\"uckelberg procedure. This corresponds to performing a time re-parametrization $t\rightarrow \tilde t = t - \pi(x,t)$ and assigning the transformation law  $\tilde\pi(x,\tilde t+\xi(\tilde t))=\pi(x,\tilde t)-\xi(\tilde t)$, such that the combination $\tilde t + \pi(x,\tilde t)$ is invariant under $\tilde t\rightarrow \tilde t + \xi(\tilde t)$. Thought of in this way one can identify the $\pi$ field as the Goldstone mode arising from the spontaneous breaking of time diffeomorphisms.
Under this transformation the time component of the metric changes as
\bea
g^{00}=\frac{\partial x^{0}}{\partial x'^\mu}\frac{\partial x^{0}}{\partial x'^\nu}g'^{\mu\nu}&=&(1+\dot\pi)^2 g'^{00} + 2(1+\dot\pi)\partial_i\pi g'^{0i} + g'^{ij}\partial_i\pi\partial_j\pi \nn \\ 
&=&  - \frac{1}{N^2} (1 + \dot \pi)^2  + 2 (1 + \dot \pi) \frac{N^i}{N^2} \partial_i \pi + a^2(t) \delta^{i j} \partial_i \pi \partial_j \pi , \label{pmetric}
\eea
where the new metric $g'^{\mu \nu}$ is the flat metric (\ref{pgauge}) conveniently expressed in the ADM parametrization (\ref{ADM-parametrization}). Therefore, the action~(\ref{eft-action-scalar-sector}) in the flat gauge takes the form:
\be \label{paction} 
\begin{split}
S [\pi] & = \int dx^3 dt \sqrt{-g} \Big[ \frac{1}{2} M_{\rm Pl}^2 R -M_{\rm Pl}^2 (3H^2(t+\pi) + \dot H(t+\pi)) \\ & + M_{\rm Pl}^2 \dot H(t+\pi)\left(  - \frac{1}{N^2} (1+\dot\pi)^2  + 2(1+\dot\pi) \frac{N^i}{N^2} \partial_i\pi  + g^{ij}\partial_i\pi\partial_j\pi \right) \\ & + \sum_{n=1}^\infty\frac{M_n^4(t+\pi)}{n!} \left( - \frac{1}{N^2} (1+\dot\pi)^2  + 2(1+\dot\pi) \frac{N^i}{N^2} \partial_i\pi  + g^{ij}\partial_i\pi\partial_j\pi + 1 \right)^n \Big] \ .
\end{split}
\ee
 Although this is quite a  complicated action when expanded out, there is a physical limit one can take where the whole situation simplifies considerably. This is the so-called decoupling limit \cite{Cheung:2007st}, in which the effects of gravity decouple from the scalar degree of freedom. From the second line of the action \eqref{paction} one can see that the leading mixing term of the scalar mode with gravity is of the form $M_{\rm Pl}^2 \dot H\dot\pi\delta g^{00}$, where $\delta g^{00} \equiv 1 / N^2 -1$. Then, after canonically normalizing the fields using $\pi_c=M_{\rm Pl} |\dot H|^{1/2} \pi/c_{\rm s}$, and $\delta g_c^{00}=M_{\rm Pl}\delta g^{00} $, this term reads $\sqrt{\epsilon} H c_{\rm s}\dot\pi_c \delta g_c^{00}$. Thus one may conclude that for energies 
 \be 
 \label{Emix} \omega \gg \sqrt{\epsilon}Hc_{\rm s} ,
 \ee 
 one can neglect such mixing terms and consider only the dynamics of the Goldstone mode. This is the analogue of the equivalence theorem \cite{Cornwall:1974km,Vayonakis:1976vz} which states that in the context of a spontaneously broken gauge symmetry, there exists an energy above which the would-be Goldstone mode decouples from the gauge field, and becomes a dynamical scalar degree of freedom. As we see from \eqref{Emix}, this limit corresponds to the slow-roll approximation in the inflationary context. This fact can also be nicely demonstrated by writing the quadratic action to first order in slow-roll.

One can compute the action $S[\pi]$ to any desired order in $\pi$.
For instance, using the ADM decomposition explicitly with $N \equiv 1 + \delta N$, we can rewrite the quadratic part of the action \eqref{paction} in the decoupling limit as
\bea
S^{(2)} &=& M_{\rm Pl}^2 \int d^3xdt a^3 \bigg[ - 3 H^2 \delta N^2 - 6 H \dot H \pi \delta N - 3 \dot H^2 \pi^2  \nn \\
&& - 2 \partial_i N^i (H \delta N + \pi \dot H)   + \dot H \frac{(\partial \pi)^2}{a^2}  - \frac{\dot H}{c_{\rm s}^2} (\delta N - \dot \pi)^2  \bigg] ,
\eea
where we have defined the speed of sound $c_{\rm s}$ through the relation
$$ \frac{1}{c_{\rm s}^2} = 1 + \frac{2 M_2^4 }{M_{\rm Pl}^2 | \dot H| } \  . $$
The linear constraint equations may be solved to give
\bea
\delta N = \epsilon H \pi , \qquad \partial_i N^i = - \frac{\epsilon }{c_{\rm s}^2} \frac{d }{d t} (H \pi).
\eea
After substituting these relations back into the action, we obtain the second-order action
\bea \label{pactionsr}
S^{(2)} &=& - M_{\rm Pl}^2 \int d^3xdt a^3 \dot H \bigg[   ( \epsilon H \pi - \dot \pi) \frac{1}{c_{\rm s}^2} ( \epsilon H \pi - \dot \pi) -   \frac{(\partial \pi)^2}{a^2}  \bigg] .
\eea
Finally, from \eqref{zofpi} truncated to first order so that $\zeta =  - H \pi $, we immediately see that
\bea
S^{(2)} &=&  M_{\rm Pl}^2 \int d^3xdt a^3 \epsilon \bigg[ \frac{ \dot\zeta^2}{c_{\rm s}^2} - \frac{(\partial \zeta)^2}{a^2}  \bigg] .
\eea
The action \eqref{pactionsr} contains a mass term for $\pi$, coming from mixing with gravity, whose form for the canonically normalized field $\pi_c$ is $\epsilon H^2 \pi_c^2 .$ Therefore, the deeper one goes in the decoupling limit, the closer to massless the $\pi$ field becomes, approaching a true Goldstone mode. Since this mass is of order the slow-roll parameters, the decoupling limit coincides with  a slow-roll expansion. In this limit, the flat gauge and particularly the EFT formalism in this gauge, can be the most useful description of inflationary perturbations. 


\section{EFT and the new physics regime} \label{EFT-new-physics} 
\setcounter{equation}{0}
In this section, we reformulate the standard EFT action (\ref{starting-action}) such that it includes a regime compatible with modified dispersion relations, which we derive,  showing explicitly that this lifts the strong coupling scale well above the UV cutoff scale $\Lambda_{\rm UV}$. This gives a generalized formalism which is weakly coupled throughout the low-energy regime and describes new physics in the form of a modified dispersion relation and scale-dependent interactions. 

\subsection{Parametrization of the new physics regime}\label{param_new_phys}

Here we show that action (\ref{starting-action}) has a natural extension in which the $ M_n^4 (1+ g^{00})^n / n!$ terms are modified in such a way that the EFT parametrizes the new physics regime. In what follows we drop the extrinsic curvature terms by setting $\bar M_n = 0$ and focus on the scalar sector of the theory. As a concrete guideline, we start by examining the extension to (\ref{starting-action}) due to the presence of heavy fields interacting with the gravitational potential $(1+ g^{00})$. The main idea is the following: terms beyond the single scalar field paradigm ($M_n = 0$) will appear as effective interaction terms resulting from the mediation of massive particle states with propagators proportional to $(M^2 - \Box)^{-1}$, where $M$ is the mass of the field being integrated out. In this context, we expect that contributions to the EFT due to these propagators in general imply the following effective $n$-point interaction term\footnote{For simplicity, we momentarily disregard gravity, allowing ourselves to drop any effects coming from the time variation of the scale factor $a(t)$. We will amend this omission immediately, in the analysis of Section~\ref{EFT-new-physics-regime}.}:
\be
\mathcal L_{\rm EFT}^{(n)} \propto   \left[ (g^{00}+1) \frac{M^2 }{ M^2 - \Box} \right]^{n-1} (g^{00}+1)  . \label{basic-modification}
\ee
At first, the extension implied by such terms seems irrelevant. Indeed, at low energies one would expect that any contribution coming from $\Box \equiv - \partial_t^2  +  \tilde\nabla^2$ acting on $(g^{00} + 1)$ will remain suppressed with respect to the mass scale $M^{2}$, hence justifying the usual expansion:
\be
\frac{1 }{ M^2 - \Box} = \frac{1}{M^2}  + \frac{\Box}{M^4} + \cdots . \label{prop-expansion}
\ee
This in turn implies that $\mathcal L_{\rm EFT}^{(n)} \propto (g^{00}+1)^{n}$ constitutes the leading order contribution to the effective action induced by heavy fields, bringing us back to the standard result (\ref{starting-action}).
However, because time translation invariance is broken, the system may find itself in a non-relativistic regime where the expansion (\ref{prop-expansion}) becomes a poor representation of the low-energy kinematics. In a non-relativistic time-dependent background, we are instead allowed to consider the case in which a hierarchy of scales appears between frequency and momenta, leading to the more general possibility:
\be
\frac{1 }{ M^2 - \Box} = \frac{1}{M^2 - \tilde\nabla^2}  - \frac{\partial_t^2 }{(M^2 - \tilde\nabla^2)^2} + \cdots .\label{prop-expansion-2}
\ee
It is clear that such an expansion is only possible in a regime where the propagation of Goldstone bosons is characterized by a non-relativistic dispersion relation $\omega(p)$ satisfying
\be
\omega^2 \ll M^2 + p^2, \label{low-energy-condition}
\ee
which in turn defines the low-energy regime. Just like in (\ref{prop-expansion}), the expansion in (\ref{prop-expansion-2}) implies that additional degrees of freedom other than the Goldstone boson ---due to the higher time derivatives--- will remain non-dynamical at low energies. In particular, the splitting implied by (\ref{low-energy-condition}) allows for physical situations where the momentum is much larger than the mass scale (i.e. $p^2 \gg M^2$) without the appearance of additional UV degrees of freedom. The end result is a low-energy effective field theory with scale-dependent $n$-point interactions of the general form:
\be
\mathcal L_{\rm EFT}^{(n)} \propto   \left[ (1+ g^{00}) \frac{M^2 }{ M^2 - \tilde\nabla^2} \right]^{n-1} (1+g^{00})  . \label{basic-modification-2}
\ee
In what follows, we analyze the direct consequence of having a low-energy EFT with contributions of the form given by (\ref{basic-modification-2}). In Section \ref{integration-1-massive-field} we present a concrete realization where (\ref{basic-modification-2}) is obtained from a well-defined system where the inflaton interacts with a single massive field. There, we also show that the massive field leading to (\ref{basic-modification-2}) becomes a Lagrange multiplier at low energies.


\subsection{EFT of the new physics regime and a modified dispersion relation} \label{EFT-new-physics-regime}
 
Following the previous discussion, we now consider the full EFT action, taking into account the $n$-point contributions of the form (\ref{basic-modification-2}) and including gravity. This leads to:
\bea \label{singlefieldEFT}
S_{\rm EFT} &=& \int d^3xdt \sqrt{-g} \bigg[ \frac{M_{\rm Pl}^2}{2} R + M_{\rm Pl}^2 \dot H g^{00} - M_{\rm Pl}^2 (3 H^2 + \dot H) \nn \\
&&  + \sum_{n = 2}^{ \infty} \frac{M_n^4 }{n!} \left[ (1+ g^{00})   \frac{M^2 }{ M^2 - \tilde\nabla^2} \right]^{n-1}  (1+g^{00}) + \cdots  \bigg] .
\eea
 We shall justify this form in Section~\ref{integration-1-massive-field}, where we deduce it in the specific example where the Goldstone boson is coupled to a single heavy field by turns of the inflationary trajectory in field space. A more general argument is given in  Appendix~\ref{several-massive-fields}.  Notice that we have constructed this action by inserting the ratio $M^2 / (M^2 - \tilde\nabla^2)$ to keep the $M_n$ parameters dimensionful. In this way, the resulting EFT theory will be completely parametrized by the $M_n^4$ coefficients and the new mass scale determined by $M$. 
As discussed in Section~\ref{Goldstone-boson-computations}, the Goldstone boson is related to $g^{00}$ of the unitary gauge by
\be
g^{00}  =  - \frac{1}{N^2} (1 + \dot \pi)^2  + 2 (1 + \dot \pi) \frac{N^i}{N^2} \partial_i \pi + a^2(t) \delta^{i j} \partial_i \pi \partial_j \pi , \label{g00-Goldstone}
\ee
where $N$ and $N^i$ are the lapse and shift functions of the ADM decomposition~(\ref{ADM-metric}) and $a(t)$ is the scale factor of the FLRW background. Then, by writing the action in terms of the Goldstone boson mode $\pi$ up to cubic order following the same steps shown in Section~\ref{Goldstone-boson-computations}, we obtain
\bea
\nonumber S^{(3)} &=& -M_{\rm Pl}^2 \int d^3xdt a^3 \dot H \bigg[   \dot \pi \bigg( 1 + \frac{2 M_2^4 }{M_{\rm Pl}^2 |  \dot H| }  \frac{M^2}{M^2 - \tilde \nabla^2} \bigg) \dot \pi -   (\tilde \nabla \pi)^2  \bigg] \\
&& +  \int d^3xdt a^3 \bigg[   2 M_2^{4}  \bigg(   \dot \pi^2 -  (\tilde \nabla \pi)^2   \bigg) \frac{M^2}{M^2 - \tilde \nabla^2} \dot \pi   -  \frac{4}{3} M_3^{4}   \bigg(  \dot \pi  \frac{M^2}{M^2 - \tilde \nabla^2}  \bigg)^2 \dot \pi    \bigg]. \label{EFT-new-physics-1}
\eea 
Notice that the standard EFT action for the Goldstone boson is recovered by taking the formal limit $M^2 \to  \infty$. In the present case, the Goldstone boson has acquired a non-trivial kinetic term with a strong scale dependence. As a consequence, the dispersion relation characterizing the free theory is
\be
\omega^2(p) =   \frac{M^2 + p^2}{ M^2 c_{\rm s}^{-2} + p^2} p^2 , \label{full-modified-dispersion}
\ee
where $c_{\rm s}$ is the speed of sound defined in the long wavelength limit as
\be \label{full-modified-cs}
\frac{1}{c_{\rm s}^2} = 1 + \frac{ 2 M_2^{4} }{ M_{\rm Pl}^2 | \dot H| }   ,
\ee
and $p \equiv k/a$ is the physical momentum. 
Recall that the expansion in (\ref{prop-expansion-2}) is valid only in the low-energy regime defined by  $ \omega^2 \ll M^2 + p^2$.
In terms of momentum, this condition is equivalent to
\be
p^2 \ll M^2 c_{\rm s}^{-2} .
\ee
In this limit the dispersion relation may be expanded as 
\be
\omega^2 (p) =   c_{\rm s}^2 p^2  + \frac{   ( 1  - c_{\rm s}^2 )   }{  M^2 c_{\rm s}^{-2}  } p^4 + \mathcal O (p^6) , \label{dispersion-cuadratic}
\ee
the term proportional to $p^6$ being always subleading.
The cutoff energy $\Lambda_{\rm UV}$ determining the validity of this expansion can be estimated by evaluating $\omega$ at the value $p =  M c_{\rm s}^{-1}$, giving
\be \label{luv}
\Lambda_{\rm UV}^2 \sim M^2 c_{\rm s}^{-2}.
\ee
Thus, $\Lambda_{\rm UV}$ represents a simultaneous cutoff scale for both momentum $p$ and energy $\omega$. Above this scale, the expansion~(\ref{prop-expansion-2})  breaks down explicitly and the system has to be UV-completed in such a way that it incorporates the states characterized by the mass scale $M c_{\rm s}^{-1}$ as new degrees of freedom. 
The value of $c_{\rm s}^2$ determines the size of the non-trivial effects due to the propagators coming from the heavy physics sector.
The quadratic piece in (\ref{dispersion-cuadratic}) dominates when $p^2 \ll M^2$, whereas the quartic term dominates when the momentum is in the range $M^2 \ll p^2 \ll   M^2 c_{\rm s}^{-2}$. The associated threshold energy scale is $\Lambda_{\rm new} \sim M c_{\rm s} $, found by evaluating $\omega^2$ at $p^2 \sim M^2$. We thus see that interesting effects related to the running of momentum in the EFT coefficients only occur for $c_{\rm s}^2 \ll 1$. We may rewrite the new physics action in a way that incorporates $c_{\rm s}$ explicitly as
\bea
S &=& -M_{\rm Pl}^2 \int d^3xdt  a^3 \dot H \bigg[    \dot \pi \left( 1 +  \Sigma (\tilde\nabla^2)  \right) \dot \pi -   (\tilde \nabla \pi)^2     +     \big[   \dot \pi^2 -  (\tilde \nabla \pi)^2   \big] \Sigma (\tilde\nabla^2) \dot \pi   \nn \\
&& -   \frac{2 M_3^{4}  c_{\rm s}^2 }{3M_2^4 (1 - c_{\rm s}^2 )   }   \, \dot \pi   \Sigma (\tilde\nabla^2) \left( \dot \pi \Sigma (\tilde\nabla^2)\dot \pi \right)    \bigg]  ,  \label{EFT-new-physics-2}
\eea
where $\Sigma (\tilde\nabla^2)$ is a differential operator given by
\be \label{Sigma-def}
\Sigma (\tilde\nabla^2) =  (1 - c_{\rm s}^{2} )  \frac{M^2 c_{\rm s}^{-2} }{M^2 - \tilde \nabla^2} .
\ee
We now see that $\Sigma (\tilde\nabla^2)$ determines the structure of interactions at both quadratic and cubic order. As the energy increases, the scale dependence of $\Sigma(\tilde\nabla^2)$ affects the strength of self interactions of $\pi$, potentially modifying the computation of $n$-point correlation functions, and any phenomenological conclusions derived from it. In particular, provided that $c_{\rm s}^2 \ll 1$, in the new physics regime one has 
\be
\Sigma (\tilde\nabla^2) \to -   \frac{  M^2 c_{\rm s}^{- 2} }{\tilde\nabla^{2}} .
\ee
Finally let us clarify that, just as for the case described by the effective theory (\ref{starting-action}), its extended version (\ref{EFT-new-physics-1}), taken on its own, provides no explicit information about the value of the UV cutoff scale $\Lambda_{\rm UV}$ at which the effective field theory breaks down. In other words, the theory (\ref{EFT-new-physics-1}) may be taken literally as it reads all the way up to momenta $p \gg M c_s^{-1}$, for which the dispersion relation (\ref{full-modified-dispersion}) becomes $\omega^2 (p) \simeq  p^2$, consistent with a Lorentz invariant spectrum of massless particles. 
However, to keep our discussion on firm physical grounds, we assume a UV cutoff $\Lambda_{\rm UV}$, consistent with there being a regime of  UV physics where the Goldstone boson interacts with one or more heavy fields.

\subsection{The strong coupling scale} \label{thestrongcouplingscale}

We now deduce an immediate consequence of the scale-dependence of $\Sigma(\tilde\nabla^2)$: 
we derive the energy scale $\Lambda_{\rm s.c.}$ at which the theory (\ref{EFT-new-physics-2}) becomes strongly coupled. 
Before proceeding, let us briefly summarize the scales that appear in the problem. These are:
\begin{itemize}
\item $\Lambda_{\rm new}$: the new physics scale which signals the change from a linear to a non-linear dispersion relation.
 
\item $\Lambda_{\rm UV}$: the UV cutoff scale given in \eqref{luv}, above which all scalar fields are dynamical and the single field effective description is no longer applicable.
 
\item $\Lambda_{\rm s.c.}$: the strong coupling scale at which a perturbative approach becomes inconsistent, resulting in a breakdown of the effective description.

\item $\Lambda_{\rm s.b.}$: the symmetry breaking scale at which time diffeomorphisms break and the effective description on a time-dependent gravitational background becomes available.
\end{itemize}
So far we have identified that the extended EFT (\ref{EFT-new-physics-2}) implies $\Lambda_{\rm new} \sim M c_{\rm s}$ and $\Lambda_{\rm UV} \sim M c_{\rm s}^{-1}$. Then, a suppressed speed of sound automatically induces the hierarchy
\be 
\Lambda_{\rm new} \ll \Lambda_{\rm UV} ,
\ee
but it tells us nothing about the relative values of $\Lambda_{\rm s.c.}$ and $\Lambda_{\rm s.b.}$ with respect to $\Lambda_{\rm UV}$, as their values strongly depend on the specific UV realization of the inflationary model at hand. Nevertheless, if the UV physics allowing for the existence of a new physics energy regime is also responsible for generating inflation, it is reasonable to expect all of these scales to be of the same order:
\be
\Lambda_{\rm UV} \sim \Lambda_{\rm s.b.} \sim \Lambda_{\rm s.c.} .
\ee
We emphasize that the present analysis is strictly valid only for inflationary models with a single scalar degree of freedom driving inflation, and that models with multiple degrees of freedom will inevitably introduce a larger set of scales into the problem.

First, following the discussion of ref.~\cite{Baumann:2011su}, it is possible to deduce that the symmetry breaking scale $\Lambda_{\rm s.b.}$, taking into account the fact that the dispersion relation scales as $\omega \sim p^2$, is given by:
\be
\Lambda_{\rm s.b.} = \left[ \frac{2 M_{\rm Pl}^2 | \dot H |}{\Lambda_{\rm UV}^4 } \right]^{2/7} \Lambda_{\rm UV}. \label{symmetry-breaking-scale}
\ee
This result allows us to see that the value of $\Lambda_{\rm s.b.}$ compared to $\Lambda_{\rm UV}$ depends on the ratio $2 M_{\rm Pl}^2 | \dot H | /  \Lambda_{\rm UV}^4$. For instance, if the UV physics in charge of modifying the low-energy dynamics of curvature perturbations is also responsible for producing inflation, it is perfectly feasible to have $ \Lambda_{\rm UV}^4 \sim 2 M_{\rm Pl}^2 | \dot H | $, implying $\Lambda_{\rm s.b.}  \sim \Lambda_{\rm UV}$. For the benefit of the present discussion, we adopt the optimistic perspective whereby $ \Lambda_{\rm s.b.} \sim \Lambda_{\rm UV} $, consistent with a large value for the slow-roll parameter $\epsilon$ 
compatible with observations (see Section~\ref{Implications-inflation}). 

We now proceed to calculate the strong coupling scale $\Lambda_{\rm s.c.}$, that is, the scale at which tree-level interactions violate unitarity.
 As discussed in ref.~\cite{Cheung:2007st}, a reduced speed of sound $c_{\rm s}^2 < 1$ inevitably introduces non-unitary self-interactions for the Goldstone boson that make the theory~(\ref{starting-action}) strongly coupled at an energy scale $\Lambda_{\rm s.c.}$ given by:
\be
\Lambda_{\rm s.c.}^4 = 4 \pi M_{\rm Pl}^2 |\dot H | c_{\rm s}^5 (1 - c_{\rm s}^2)^{-1} . \label{standard-sc}
\ee
However, this result is strictly valid only for the case in which $\omega^2 = c_{\rm s}^2 p^2$, characteristic of the standard EFT picture. In fact, in ref.~\cite{Baumann:2011su}  it was found that a modification of the dispersion relation will generally alleviate the strong coupling problem by making $\Lambda_{\rm s.c.}$ larger than the  value of (\ref{standard-sc}). Nevertheless, in that work a general analysis incorporating the scale dependence of self-interactions consistent with the modification of the dispersion relation was not taken into account. In what follows we incorporate this aspect into the analysis of strong coupling and show that the conditions for the theory to remain weakly coupled are satisfied all the way up to an energy scale of the same order as, or larger than, the natural cutoff $\Lambda_{\rm UV} = M c_{\rm s}^{-1}$ of the new physics regime. 

To proceed, we will follow the analysis of ref.~\cite{Baumann:2011su} closely. First, by normalizing the Goldstone boson as $\pi_n = ( 2 M_{\rm Pl}^2 \epsilon H^2  )^{1/2} \pi$ the quadratic part of the extended EFT action (\ref{EFT-new-physics-2}) may be written as
\bea
S &=& \frac{1}{2}  \int d^3xdt  \bigg[    \dot \pi_n \left(   \frac{   M^2 c_{\rm s}^{-2} -  \nabla^2  }{M^2 -  \nabla^2}  \right) \dot \pi_n -   ( \nabla \pi_n)^2   \bigg] .
\eea
Notice that we have fixed $a=1$ for the sake of simplicity, assuming that our discussion involves processes at energy scales much larger than $H$. Now, the conjugate momentum $P_{\pi} \equiv \partial \mathcal L / \partial \dot \pi_n $ of this free field theory is given by
\be
P_{\pi}  = \frac{M^2 c_{\rm s}^{-2}  - \nabla^2}{   M^2 -   \nabla^2} \dot \pi_n .
\ee
This implies that the commutation relation $[\pi_n , P_{\pi}] = i \delta$ reads as
\be
\left[ \pi_n ({\bf x}_1)  \, , \frac{M^2 c_{\rm s}^{-2} -  \nabla^2_2}{   M^2 -   \nabla^2_2} \dot \pi_n  ({\bf x}_2) \right] = i \delta ({\bf x}_1 - {\bf x}_2 ) , \label{commutation-new}
\ee
where $\nabla^2_2$ stands for a Laplacian operator written in terms of the coordinate ${\bf x}_2$. Another way of writing this expression is:
\be
\left[ \pi_n  ({\bf x}_1) \, ,  \dot \pi_n  ({\bf x}_2) \right] = i  \frac{M^2 -  \nabla^2_2}{   M^2 c_{\rm s}^{-2} -   \nabla^2_2} \delta ({\bf x}_1 - {\bf x}_2 ) .
\ee
Then, to satisfy these commutation relations, we may consider the quantization of the free field $\pi_n (x)$ in terms of creation and annihilation operators $\hat a^{\dag}({\bf k})$ and $\hat a({\bf k})$ satisfying \\ $[\hat a({\bf k}_1) , \hat a^{\dag}({\bf k}_2) ] = \delta({\bf k}_1 - {\bf k}_2)$. We find
\be
\pi_n (x) = \frac{1}{(2 \pi)^{3/2}} \int d^3 p \left[ \pi_n (p) \hat a({\bf p}) e^{- i \omega t +  i {\bf p } \cdot {\bf x} }  + \pi_n (p)^{*} \hat a^{\dag}({\bf p}) e^{+ i \omega t -  i {\bf p } \cdot {\bf x} }   \right] ,
\ee
where $\pi_n (p)$ corresponds to the field amplitude in Fourier space, given by:
\be
\pi_n (p) = \sqrt{ \frac{M^2 +p^2}{   M^2 c_{\rm s}^{-2}  +  p^2 } } \frac{1}{\sqrt{2\omega(p)}} = \sqrt{\frac{\omega(p)}{2 p^2}} . \label{new-amplitude}
\ee
Notice that due to the modified commutation relation (\ref{commutation-new}), the functional form of $\pi_n (p)$ differs substantially from the standard Minkowskian result $1 / \sqrt{2 \omega}$. Nevertheless, it is possible to see that when $p \ll \Lambda_{\rm new} = M c_{\rm s}$ we recover back the standard amplitude $1/\sqrt{2 c_{\rm s} p}$ after canonically normalizing $\pi_c = \pi_n / c_{\rm s}$. Apart from this modification, the quantization of the quadratic action proceeds in the usual way. 

Let us now move on to consider the interacting part of the theory. Notice that the relevant quartic interaction due to $M_2^4$, coming from the non-linear self-interactions in the EFT is given by:\footnote{Another relevant interaction that could contribute to this analysis is the one proportional to $\mathcal L_{\rm int } \propto \dot \pi^2  \Sigma (\tilde\nabla^2)  \dot \pi^2 $. However, in the new physics regime one has $\omega^4 \ll p^4$, implying that this interaction will be substantially suppressed compared to (\ref{strong-coupling-interaction}).}
\be
\mathcal L_{\rm int} =  \frac{ (1 - c_{\rm s}^{2} )}{16 M_{\rm Pl}^2 \epsilon H^2}  ( \nabla \pi_n)^2     \frac{M^2 c_s^{-2} }{M^2 -  \nabla^2}  ( \nabla \pi_n)^2 , \label{strong-coupling-interaction}
\ee
after taking into account the normalization $\pi_n = ( 2 M_{\rm Pl}^2 \epsilon H^2  )^{1/2} \pi$.
Then, we can analyze the effect of this interaction on the tree-level scattering of two $\pi$ fields into two final $\pi$'s in the center of mass reference frame. The main point to be kept in mind when performing this computation is that the new amplitude (\ref{new-amplitude}) for the quantized Goldstone boson field implies that each external leg of the relevant diagram will come with an additional factor 
\be
\sqrt{ \frac{M^2 +p_i^2}{   M^2 c_{\rm s}^{-2} +   p_i^2 } } ,
\ee
where $p_i$ is the momentum carried by the particle represented by the $i$th external leg of the diagram. After a straightforward computation,
we find that the scattering amplitude of this interaction is given by:
\bea 
\mathcal A (p_1 , p_2 \to p_3 , p_4) &=& \frac{(1-c_{\rm s}^2) c_{\rm s}^{-2} p^4}{2 M_{\rm Pl}^2 | \dot H | } \left( \frac{M^2 +p^2}{   M^2 c_{\rm s}^{-2} +   p^2 } \right)^2 \nn \\
&& \qquad \qquad \times \bigg[ 1 + \frac{M^2 \cos^2 \theta}{M^2 + 2 p^2 (1 - \cos \theta)} + \frac{M^2 \cos^2 \theta}{M^2 + 2 p^2 (1 + \cos \theta)} \bigg], \label{scatt-amplitude}
\eea
where $\theta$ is the angle of scattered particles with respect to the impact axis. By recalling that $(M^2 - \nabla^2)^{-1}$ can be interpreted as the propagator of a heavy field, the first, second and third terms in the square bracket of (\ref{scatt-amplitude}) may be thought as contributions coming from the interchange of a heavy boson through the $s$, $t$ and $u$ channels respectively. The previous result may be expressed by the partial wave expansion
\be
\mathcal A (p_1 , p_2 \to p_3 , p_4)=  16 \pi \left( \frac{\partial \omega}{\partial p} \frac{\omega^2}{p^2} \right) \sum_\ell (2 \ell + 1) P_{\ell} (\cos \theta) a_\ell,
\ee
where the $P_{\ell} (\cos \theta)$ are Legendre polynomials. Then the lowest order coefficient $a_0$ is given by:
\bea
a_0 &=& \frac{(1-c_{\rm s}^2) c_{\rm s}^{-2} p^4}{32 \pi M_{\rm Pl}^2 | \dot H | }  \left( \frac{\partial p}{\partial \omega} \frac{p^2}{\omega^2} \right)  \left( \frac{M^2 +p^2}{   M^2 c_{\rm s}^{-2} +   p^2 } \right)^2\int_{-1}^{1} d\cos \theta \left( 1 + \frac{2 ( M^2 + 2 p^2) M^2 \cos^2 \theta}{( M^2 + 2 p^2)^2  - 4 p^4 \cos^2 \theta } \right) \nn \\
&=& \frac{(1-c_{\rm s}^2) c_{\rm s}^{-2} p^4}{16 \pi M_{\rm Pl}^2 | \dot H | }  \left( \frac{\partial p}{\partial \omega} \frac{p^2}{\omega^2} \right) \left( \frac{M^2 +p^2}{   M^2 c_{\rm s}^{-2} +   p^2 } \right)^2 \nn \\ && \qquad \qquad \qquad \qquad \qquad \quad \times \bigg[ 1 + \frac{ M^2 ( M^2 + 2 p^2)}{2 p^4} \left(  \frac{ M^2 + 2 p^2}{4 p^2} \log \Big(1 + \frac{4 p^2 }{ M^2}  \Big) - 1 \right)  \bigg]. \qquad \label{a-0-result-1}
\eea
In order to preserve the unitarity of the tree-level scattering process under consideration, the optical theorem leads to the constraint $a_\ell + a_{\ell}^{*} \leqslant 1$. Our main interest is to assess the unitarity of the EFT above $\Lambda_{\rm new}$, where the dispersion relation has the form:
\be
\omega^2 \simeq \frac{p^4}{M^2 c_{\rm s}^{-2}}  .
\ee
In this regime, the second term in the square bracket of (\ref{a-0-result-1}) becomes negligible, leading to the result 
\be
a_0 \simeq \frac{  (M c_{\rm s}^{-1})^{3/2 } \omega^{5/2}}{32 \pi M_{\rm Pl}^2 | \dot H | c_{\rm s}^2 } .   
\ee
Then, using the constraint $a_\ell + a_{\ell}^{*} \leqslant 1$, i.e. ${\rm Re}(a_\ell) < \frac{1}{2}$,  for the particular case $\ell = 0$, we find that the theory remains weakly coupled as long as
\be
\omega^{5/2} < 8 \pi c_{\rm s}^2 \left[ \frac{ \Lambda_{\rm s.b.} }{ \Lambda_{\rm UV}} \right]^{7/2} \Lambda_{\rm UV}^{5/2} .
\ee
From this result we deduce that the strong coupling scale is given by:
\be \label{strong-coupling-scale}
\Lambda_{\rm s.c.} = (8 \pi c_{\rm s}^2)^{2/5}  \left[ \frac{ \Lambda_{\rm s.b.} }{ \Lambda_{\rm UV}} \right]^{7/5} \Lambda_{\rm UV} ,
\ee
where $\Lambda_{\rm s.b.}$ is the symmetry breaking scale (\ref{symmetry-breaking-scale}). Equation~(\ref{strong-coupling-scale}) admits a variety of situations depending on the values of the scales $\Lambda_{\rm new}$, $\Lambda_{\rm UV}$ and $\Lambda_{\rm s.b.}$.  For instance, if we take $\Lambda_{\rm s.b.} \sim \Lambda_{\rm UV}$ and  $c_{\rm s}^2 = \Lambda_{\rm new} / \Lambda_{\rm UV} \sim 10^{-4}$, then $(8 \pi c_{\rm s}^2)^{2/5} \sim 0.1$ and the value of $\Lambda_{\rm s.c.}$ is found to be of order $\Lambda_{\rm UV}$. However, if $\Lambda_{\rm s.b.} \gg \Lambda_{\rm UV}$, then one can have models with $c_{\rm s}^2 \ll 10^{-4}$ and still satisfy $\Lambda_{\rm s.c. } \sim \Lambda_{\rm UV} $. Thus we see that the non-trivial modifications characterizing the new physics regime $M^2 \ll p^2 \ll M^2 c_{\rm s}^{-2}$ imply that the interactions of the theory scale differently with energy, changing significantly the value at which the EFT becomes strongly coupled.


\section{The new physics regime from heavy fields} \label{integration-1-massive-field}
\setcounter{equation}{0}
Actions containing an arbitrary number of inverse differential operators, such as (\ref{singlefieldEFT}), are known to be non-local, potentially suffering from classical instabilities and the appearance of ghosts at the quantum level, when the non-local terms can be written as the limit of a sequence of higher-derivative terms. Indeed, Ostrogradski found that theories which depend non-trivially on more than one time derivative ({\it i.e.} in such a way that the higher derivatives cannot be removed by integrating by parts) are unstable, with their Hamiltonians unbounded from below~\cite{Ostrogradski, Woodard:2006nt}. Upon quantization the instability persists, manifested by the appearance of negative norm states or ghosts, which in turn translates into loss of unitarity.
However, when the theory in question corresponds to an effective field theory derived from a local theory by integrating out one or more fundamental dynamical variables, we do not encounter these problems. In such a case, because the original theory is local, it is not valid to consider the resulting non-local terms as limits of higher-derivative terms \cite{Eliezer:1989cr, Bennett:1997wj}, implying that there are no problems either with instabilities or ghosts, as long as the theory remains within its domain of validity.

In what follows we explicitly relate the non-local form of action (\ref{singlefieldEFT}) to the presence of additional degrees of freedom that become operative at high energies. We show that the theory at hand becomes ill defined only if one insists in assuming its validity at energies of order $M c_{\rm s}^{-1}$, where a second degree of freedom inevitably becomes excited. The result is that at low energies the theory (\ref{singlefieldEFT}) is safe from any pathology related to non-locality.

\subsection{Integration of a single massive field} \label{singleheavyfieldexample}

As an illustrative example, let us study the specific case of the EFT obtained by integrating out a single massive field parametrizing deviations from the trajectory in field space \cite{Achucarro:2010jv, Achucarro:2010da, Cespedes:2012hu, Achucarro:2012sm, Achucarro:2012yr}.\footnote{The potentially large influence that heavy fields could have on the low-energy dynamics of inflation was first emphasized by Tolley and Wyman in ref.~\cite{Tolley:2009fg}. For other recent approaches studying the effects of heavy fields on the low-energy dynamics of inflation, see for instance refs.~\cite{Cremonini:2010sv, Jackson:2010cw, Cremonini:2010ua, Jackson:2011qg, Shiu:2011qw, Avgoustidis:2012yc, Pi:2012gf, Gao:2012uq, Burgess:2012dz}.} This gives rise to terms with insertions of the form $\frac{M^2}{M^2 - \nabla^2}$, providing a physical motivation for the modified action (\ref{modified-action-2}).  In the more general case, one obtains a more complicated function of the momenta and couplings instead of the insertion $\frac{M^2}{M^2 - \nabla^2}$, and a more complicated dispersion relation (see Appendix \ref{several-massive-fields} for the relevant analysis).  In unitary gauge, the quadratic action for a single heavy field fluctuation $\mathcal F$ coupled to the inflaton, is given by
\bea
S_{\mathcal F} &=&  \frac{1}{2}  \int \!\!  d^3xdt a^3  \bigg\{     \dot { \mathcal F}^2 - (\nabla \mathcal F)^2  -  M^2  \mathcal F^2  - 2 \dot \theta \dot \phi_0 \mathcal F  (  g^{00}  + 1 )  
 -  \dot \theta^2 \mathcal F^2 (  g^{00}  + 1 )    \bigg\}  , \label{heavy-field-unitary-gauge}
\eea
where $\dot \theta$ is the angular velocity characterizing the turns of the multi-field trajectory in the scalar field target space, and $\dot \phi_0$ is the rapidity of the background scalar field. Then, the equation of motion of the heavy field reads
\be
\ddot { \mathcal F} +3H \dot {\mathcal F}  + \left(M^2 -  \nabla^2  +  \dot \theta^2 (g^{00} + 1)\right)  {\mathcal F} =  -  \dot \theta \dot \phi_0  (g^{00} + 1). \label{equation-F}
\ee
We are interested in studying the low-energy regime of the system, where the second-order time variation of the heavy field $\ddot {\mathcal F}$ is subleading with respect to the term $( M^2 - \nabla^2 ) \mathcal F$. As shown in ref.~\cite{Cespedes:2012hu}, in order to integrate out $\mathcal F$ it is also important to assume the adiabaticity condition $ | \ddot \theta / \dot \theta | \ll M$, which ensures that the background dynamics of the turning trajectory are consistent with the condition $\omega^2 \ll M^2 + p^2 $. If this is granted, we may simply disregard the kinetic term and solve for $\mathcal F$ by rewriting the equation of motion (\ref{equation-F}) as
\be
 \left[1  +  (g^{00} + 1) \frac{ \dot \theta^2}{M^2 -  \nabla^2} \right] (M^2 -  \nabla^2 ) {\mathcal F} =  -  \dot \theta \dot \phi_0  (g^{00} + 1). \label{equation-F-2}
\ee
This implies that the heavy field $\mathcal F$ may be treated as a Lagrange multiplier, allowing us to explicitly write it in terms of $g^{00} + 1$ as
\bea
  {\mathcal F}   & =&  - \frac{ \dot \theta \dot \phi_0 }{M^2 -  \nabla^2  +  \dot \theta^2 (g^{00} + 1) } (g^{00} + 1) \nn \\
  &= &  -  \dot \theta \dot \phi_0  \frac{1 }{M^2 -  \nabla^2} \left[1  +  (g^{00} + 1) \frac{ \dot \theta^2}{M^2 -  \nabla^2} \right]^{-1} (g^{00} + 1) \nn \\
  &=&    - \dot \theta \dot \phi_0 \frac{1}{M^2 -  \nabla^2} \sum_{n=0}^\infty (-1)^{n} \left[ (g^{00} + 1) \frac{ \dot \theta^2}{M^2 -  \nabla^2} \right]^n (g^{00} + 1) ,\label{solution-F-g00}
\eea
where in the last step we made use of the formal expansion $(1 + x)^{-1} = \sum_n (-1)^n x^n$, valid for $|x| < 1$.
Notice that neglecting the kinetic term at the level of the equations of motion is equivalent to having dropped them in the action. Then, disregarding the kinetic term for $\mathcal F$ in the action and inserting (\ref{solution-F-g00}), we recover the contribution to the EFT for the Goldstone boson coming from the heavy field:
\be
S_{\mathcal F} =  - M_{\rm Pl}^2  \int \!\!  d^3xdt a^3  \dot H   \sum_{n=2}^{\infty}  (-1)^n  \bigg[  (g^{00} + 1)   \frac{  \dot \theta^2 }{M^2 -  \nabla^2   }  \bigg]^{n-1}  (g^{00} + 1)  , \label{resulting-action-F}
\ee
where we have used the background equation $\dot \phi_0^2 = - 2 \dot H M_{\rm Pl}^2 $. Assuming that  there are no additional sources of deviations from standard single field inflation other than the heavy field $\mathcal F$, it is straightforward to deduce that the speed of sound is given by
\be
\frac{1}{c_{\rm s}^2} = 1 + \frac{4 \dot \theta^2}{M^2}, \label{speed-of-sound-single-heavy-field}
\ee
implying that the $M_n^4$ coefficients of the EFT action (\ref{starting-action}) may be written as~\cite{Achucarro:2012sm}
\be
M_n^4 = (-1)^{n} |\dot H | M_{\rm Pl}^2 n!  \left( \frac{1 - c_{\rm s}^2}{4 c_{\rm s}^2}  \right)^{n-1} .
\ee
Thus, we have obtained the new physics EFT action (\ref{singlefieldEFT}) by integrating out a single heavy field. In this case the parameters acquire a specific dependence on the background quantity $\dot \theta$ which parametrizes the UV-complete theory (\ref{heavy-field-unitary-gauge}).  Different parameters are obtained in the general case, where one may have coupling to more than one heavy field (see Appendix \ref{several-massive-fields}). 
It is important to recognize that at low energies the massive scalar field $\mathcal F$ has no dynamics, in the sense that its value is completely determined by the source $ -  \dot \theta \dot \phi_0  (g^{00} + 1)$ at the right hand side of (\ref{equation-F-2}). In other words, the heavy field $\mathcal F$ plays the role of a Lagrange multiplier, carrying with it the scale dependence implied by the $\nabla^2$ operator.

\subsection{On the role of higher-order time derivatives} \label{UVdof}

We are now in a position to take a closer look at the role played by higher-order time derivatives in the low-energy effective field theory.
As already argued, the expansion (\ref{prop-expansion-2}) is only possible if Lorentz invariance is broken, which in the present context is a consequence of the broken time translation invariance induced by the background. In Fourier space eq.~(\ref{prop-expansion-2}) may be expressed as
\be
 \frac{1 }{ M^2 + p^2 - \omega^2} = \frac{1}{M^2  + p^2}  + \frac{ \omega^2 }{(M^2  + p^2)^2} + \cdots .\label{prop-expansion-momentum}
\ee
Here we wish to address the relevance of $\omega^2$ in this expansion, and its effect on the low-energy dynamics. To proceed, let us consider the {\it quadratic} action for $\pi$ obtained by integrating out the heavy field $\mathcal F$ of the previous section, but this time keeping the time derivative $\partial_t$ to all orders. Then, the effective action obtained for $\pi$ is found to be
\be
S_\pi^{(2)} = -M_{\rm Pl}^2 \int d^3xdt  a^3 \dot H \left[ \dot \pi \left(  1 +  \frac{  4 \dot \theta^2  }{ M^2 - \Box }  \right) \dot \pi -  (\nabla \pi)^2 \right] . \label{S-2-pi}
\ee
From this expression, it is possible to read off the propagator $D(p^2)$ of the low-energy Goldstone boson in momentum space, which is found to be given by:
\be
D(p^2) \propto \frac{1}{ \Gamma(p^2)} , \qquad  \Gamma(p^2) = p^2 - \omega^2  -  \frac{  4 \dot \theta^2  \omega^2 }{ M^2 + p^2 - \omega^2  }  .
\ee
This propagator has two poles, at values $\omega_{+}^2$ and $\omega_{-}^2$ determined by the following expression:
\begin{equation}\label{frequencies}
 \omega^{2}_{\pm} = \frac{M^2 }{2 c_{\rm s}^2} +  p^2   \pm \frac{M^2 }{2  c_{\rm s}^{2}}  \sqrt{  1 +  \frac{4 p^2 ( 1 - c_{\rm s}^{2}) }{M^2 c_{\rm s}^{-2}}   } \, . 
\end{equation}
A particle state characterized by a propagator with two or more poles is condemned to include ghosts in its spectrum~\cite{Biswas:2005qr, Barnaby:2007ve}, in close connection with our discussion on non-locality at the beginning of this section.\footnote{Strictly, this is only true for the case of an analytic generatrix, see \cite{Barnaby:2007ve}, so that exceptions such as that in \cite{Deser:2007jk} are possible. We thank Neil Barnaby for pointing this out to us.} However, if we restrict the theory to momenta $p \ll M c_{\rm s}^{-1}$, corresponding to the low-energy regime, one finds that $\omega_{+}$ and $\omega_{-}$ are well approximated by
\bea
\omega_+^2 (p) &=& M^2 c_{\rm s}^{-2} + \mathcal O (p^2) ,  \\
\omega_-^2 (p) &=&  c_{\rm s}^2 p^2 + \frac{(1 - c_{\rm s}^2)^2 }{M^2 c_{\rm s}^{-2}} p^4 + \mathcal O (p^6) , \label{low-energy-frequency-resumed-theory}
\eea
where $\mathcal O (p^2)$ and $\mathcal O (p^6)$ denote subleading higher-order terms. We thus see that, as long as we focus on low-energy processes for which $p \ll M c_{\rm s}^{-1}$ and $\omega \ll M c_{\rm s}^{-1}$, intermediate particle states are characterized by well-defined propagators (away from the dangerous pole $\omega_+$, which has a negative residue) and the effective field theory (\ref{S-2-pi}) remains ghost free. Moreover we see that there is a transition scale $\Lambda_{\rm new}$ within the low-energy regime at which the dispersion  relation $\omega_- (p)$ changes from linear to non-linear. For $c_{\rm s} \ll 1$, this roughly happens at $p =   M$, for which the first and second terms in (\ref{low-energy-frequency-resumed-theory}) compete, giving us back 
\be 
\Lambda_{\rm new} \simeq   M    c_{\rm s} .
\ee
Interestingly, (\ref{low-energy-frequency-resumed-theory}) does not coincide with our dispersion relation (\ref{dispersion-cuadratic}) computed without taking into account the time derivative $\partial_t$ to all orders. The difference between (\ref{low-energy-frequency-resumed-theory})  and (\ref{dispersion-cuadratic}) is a factor $(1 - c_{\rm s}^2)$ in front of the quartic piece of the expansion. Nevertheless, because this term is only relevant for $c_{\rm s}^2 \ll 1$, we see that the difference between these two expressions is only marginal, justifying the approximation by which one drops higher-order time derivatives.

To further appreciate the result above, we may consider again the full model (\ref{heavy-field-unitary-gauge}) coupling the scalar mode $(g^{00} + 1)$ to the heavy field $\mathcal F$, but this time taking into account the dynamics of the heavy field. In the short wavelength regime, where the role of the Hubble constant $H$ may be disregarded, the linear equations of motion for both the heavy field $\mathcal F$ and the Goldstone boson $\pi$ are given by
\bea
\ddot \pi  - \nabla^2 \pi  &=&  -  2 \frac{ \dot \theta }{\dot \phi_0}  \dot {\mathcal F} ,  \label{eq-cuad-pi-F-1} \\
\ddot {\mathcal F} - \nabla^2 \mathcal F  + M^2  \mathcal F &=&  + 2 \dot \phi_0 \dot \theta \dot \pi . \label{eq-cuad-pi-F-2}
\eea
This pair of fields is non-trivially coupled, implying that the solutions correspond to a linear combination of two modes, hereby denoted $+$ and $-$, in the form~\cite{Achucarro:2010jv, Achucarro:2012yr}
\begin{eqnarray}\label{solution-pi-F}
\pi = & \pi_+ e^{i \omega_+ t}  + \pi_- e^{i \omega_- t} \, , 
\nonumber\\
\mathcal F = &   \mathcal F_+ e^{i \omega_+ t}  + \mathcal F_- e^{i \omega_- t} \, ,  
\end{eqnarray}
where the two frequencies $\omega_{-}$ and $\omega_{+}$ are precisely those given by the expressions (\ref{frequencies}). 
The pairs $( \pi_- , \mathcal F_-)$ and $( \pi_+ , \mathcal F_+)$ represent the amplitudes of low and high frequency modes respectively. Due to the equations of motion (\ref{eq-cuad-pi-F-1}) and (\ref{eq-cuad-pi-F-2}) they satisfy the following algebraic relations
\begin{equation}\label{amplitudes}
\mathcal F_-  =    \frac{  2 i \dot \theta  \dot \phi_0 \omega_- }{ M^2 + p^2  - \omega_-^2 } \pi_-  \, , \quad  \pi_+  =  \frac{1}{\dot \phi_0} \frac{2 i \dot \theta \omega_+}{\omega_+^2 - p^2} \mathcal F_+ \, .
\end{equation}
Notice that in the limit $\dot \theta \to 0$ we recover the usual situation whereby $\mathcal F_- =  \pi_+  = 0$, and only the modes $ \pi_-$ and $\mathcal F_+$ contribute to each field.
As discussed in detail in~\cite{Achucarro:2012yr}, at tree-level, integrating out the heavy field corresponds to truncating the heavy mode of frequency $\omega_+$. This is equivalent to disregarding $\pi_+, \mathcal F_+$ and keeping the low frequency modes in the solution $\pi =  \pi_- e^{i \omega_- t} $, and $\mathcal F =    \mathcal F_- e^{i \omega_- t}$.
Of course, this step is only consistent if there is a hierarchy of frequencies $\omega_{-}^2 \ll \omega_{+}^2$, which  from (\ref{frequencies}) necessarily implies
\begin{equation}\label{range-low-energy-solution}
p^2  \ll M^2 c_{\rm s}^{-2} \, .
\end{equation}
This corresponds to the low-energy regime, and coincides with our previous criterion for avoiding ghosts at the effective field theory level. This result is twofold: First it shows explicitly how the appearance of ghosts at the effective field theory level is directly related to the appearance of heavy degrees of freedom in the full UV-complete theory and, in addition, it provides a physical explanation of the appearance of the scale $\Lambda_{\rm new}$. Upon integrating out the heavy field $\mathcal{F}$,  the dynamics of the light field $\pi$ may be thought as those corresponding to the propagation of fluctuations in a medium. From \eqref{low-energy-frequency-resumed-theory} we see that $\omega_-(p)$ has the following form in the new physics regime $\Lambda_{\rm new} \ll \omega_- \ll \Lambda_{\rm UV}$:
\be 
\label{omega_nonlinear} 
\omega_-\simeq \dfrac{p^2}{\Lambda_{\rm UV}}  + \frac{1}{2}\Lambda_{\rm new}.
\ee 
This is a Schr\"odinger dispersion relation with a mass gap $\Lambda_{\rm new}/2$ which corresponds to the scale where particle-like excitations with a non-linear dispersion relation start dominating over phonon-like ones with a definite speed of sound. Note that the effective mass of the excitation is set by the UV physics, which is responsible for the lowering of the propagation speed. Such ``gapped" Hamiltonians are familiar from many condensed matter systems, such as  super-  or semi-conductors for example.


\section{Implications for inflation} \label{Implications-inflation}
\setcounter{equation}{0}

We now discuss the observational consequences of the non-trivial modifications emerging from the extended EFT of inflation (\ref{modified-action-2}). We are particularly interested in the  distribution of curvature perturbations arising from modes which cross the Hubble scale ($\omega^2 = H^2$) within the new physics regime 
\be \label{p-range}
M^2 \ll p^2 \ll M^2 c_{\rm s}^{-2} .
\ee
Given that in this range $\omega$ is dominated by the quadratic piece ($\omega \propto p^2$), we see that in terms of $H$ the regime we are interested in is characterized by the condition
\be \label{nprangeH}
M^2 c_{\rm s}^2 \ll H^2 \ll M^2 c_{\rm s}^{-2} .
\ee
As we shall see, the observables computed in this situation do not depend on $c_{\rm s}$ directly ---as opposed to the standard case--- but instead depend on the combination $\Lambda_{\rm UV} \equiv M c_{\rm s}^{-1}$. This result changes completely the interpretation of cosmological observations as they relate to theoretical inflationary parameters such as $H$ and $\epsilon$.

\subsection{Power spectrum}
First, let us consider the derivation of the power spectrum. To proceed, we may consider the quadratic part of the action (\ref{EFT-new-physics-2}) which, for the normalized field $\pi_n = ( 2 M_{\rm Pl}^2 \epsilon H^2  )^{1/2} \pi$, reads
\bea
S &=& \frac{1}{2}  \int d^3x dt a^3 \bigg[   \dot \pi_n \left(   \frac{   M^2 c_{\rm s}^{-2} - \tilde \nabla^2  }{M^2 - \tilde \nabla^2}  \right) \dot \pi_n -   (\tilde \nabla \pi_n)^2   \bigg] ,
\eea
where $\tilde \nabla^2 \equiv \nabla^2 / a^2$. The equation of motion for $\pi_n(k)$ in momentum space is given by
\be
\ddot \pi_n + 3 H \dot \pi_n + \frac{2 H (1-c_{\rm s}^2) M^2 k^2 / a^2}{(M^2 + k^2/a^2) (M^2 + c_{\rm s}^2 k^2/a^2)} \dot \pi_n + \omega^2 \pi_n = 0 , \label{eq-motion-pi-modified}
\ee
where $k$ corresponds to co-moving momentum, and $\omega^2$ is given by eq.~(\ref{full-modified-dispersion}) with $p = k/a$. It may be seen that the third term in (\ref{eq-motion-pi-modified}) consists of a non-trivial contribution due to the scale-dependent modification of the kinetic term, whereas the fourth term contains information about the dispersion relation $\omega = \omega(p)$.\footnote{The extrinsic curvature terms appearing in action (\ref{starting-action}) do imply a modified dispersion relation, but do not generate an equation of motion such as~(\ref{eq-motion-pi-modified}). The role of these extrinsic curvature terms on inflationary observables was studied in detail in refs.~\cite{Bartolo:2010bj, Bartolo:2010di, Bartolo:2010im}.} { The equation of motion \eqref{eq-motion-pi-modified} cannot be solved analytically in full generality due to the time dependence of the coefficients through the scale factor $a(t)$. In order to obtain interesting results we will therefore assume that Hubble crossing takes place in the non-linear dispersion regime and keep the leading terms in the expansion.
Then, using the condition (\ref{p-range}), the problem simplifies to
\be
\ddot \pi_n + 5 H \dot \pi_n + \frac{k^4}{a^4 \Lambda_{\rm UV}^2} \pi_n = 0 .  \label{eq-motion-pi}
\ee
Inspection of this equation allows us to see that Hubble crossing happens at $p^2 = k^2 / a^2= H \Lambda_{\rm UV}$, consistent with the condition $\omega = H$ as long as we take the modified dispersion relation (\ref{dispersion-cuadratic}). The solution to the equation of motion (\ref{eq-motion-pi}) can be expressed in terms of the Hankel function of the first kind $ H_{5/4}^{(1)} (x)$, as~\cite{Baumann:2011su}
\be \label{pi-solution}
\pi (\tau) = \frac{H^2 \tau^2}{( 2 M_{\rm Pl}^2 \epsilon H^2  )^{1/2}} \sqrt{\frac{\pi}{8}} \frac{k}{\Lambda_{\rm UV}} \sqrt{- \tau} H_{5/4}^{(1)} (x) , \qquad x = \frac{H}{2 \Lambda_{\rm UV} } k^2 \tau^2   ,
\ee
where $\tau = - ( H a)^{-1}$ is the usual conformal time. To obtain this solution one imposes the commutation relation (\ref{commutation-new}) and chooses initial conditions such that at sub-horizon scales only positive frequency modes contribute to the propagating modes. Technically, for this procedure to be reliable, we must assume that these initial conditions are imposed within the low-energy regime $\omega^2 \ll \Lambda_{\rm UV}^2$. Taking the super-horizon limit of \eqref{pi-solution} and using the relation $\zeta = -H \pi$, we then obtain the power spectrum 
\be
\mathcal P_{\zeta} = \frac{k^3}{2 \pi^2} |\zeta_k |^2  = \frac{\Gamma^2(5/4) }{\pi^3} \frac{H^2}{M_{\rm Pl}^2 \epsilon} \sqrt{\frac{\Lambda_{\rm UV}}{H}},
\ee
where $\Gamma(5/4) \simeq 0.91$ is the usual Gamma function evaluated at $5/4$.
As mentioned earlier, the power spectrum does not depend directly on the value of $c_{\rm s}$. Because $\Lambda_{\rm UV} \gg H$ we see that the amplitude of the power spectrum is greatly enhanced (by a factor of $c_s \sqrt{\Lambda_{\rm UV}{H}}$) compared to the standard result found in single field slow-roll inflation. As a consequence , conclusions about the value of $\epsilon$ from observations of the power spectrum will change compared to the standard case.  
The normalization of the power spectrum implied by WMAP7 is given by $\mathcal P_{\zeta} = 2.42 \times 10^{-9}$~\cite{Komatsu:2010fb}, implying the following relation among the parameters determining the power spectrum
\be
\frac{H^2}{M_{\rm Pl}^2 \epsilon} \sqrt{\frac{\Lambda_{\rm UV}}{H}} \sim 9 \times 10^{-8} .
\ee
In terms of the symmetry breaking scale (\ref{symmetry-breaking-scale}) introduced earlier, the previous relation reads 
\be
 \Lambda_{\rm s.b.}^2 /  H^2  \sim 10^{4} . \label{power-spectrum-symmetry-breaking}
\ee
If $\Lambda_{\rm UV} \sim \Lambda_{\rm s.b.}$ the previous relation implies a large hierarchy of scales between the Hubble horizon and the scales involved in the ultraviolet physics.
Furthermore, given that we are considering only a modification of the scalar sector of the theory, we may safely assume that tensor perturbations remain unmodified by the presumed ultraviolet physics involved in the present analysis. This implies that the tensor-to-scalar ratio $r$ predicted for this class of EFTs has the form:
\be
r = \frac{2 \pi \epsilon}{\Gamma^2(5/4)} \sqrt{\frac{H}{\Lambda_{\rm UV}} } .
\ee
Future CMB experiments could constrain the tensor-to-scalar ratio to $r < 0.01$~\cite{Araujo:2012yh}, implying a constraint on the parameters of the form $  \epsilon \sqrt{ H / \Lambda_{\rm UV} }  \lsim 10^{-3}$. Then, in the particular case in which $\Lambda_{\rm UV} \sim \Lambda_{\rm s.b.}$, we may use (\ref{power-spectrum-symmetry-breaking}) to deduce $\epsilon <\lsim10^{-2}$, which would constitute a weaker bound that the one encountered in single field slow-roll inflation ($\epsilon < 6 \times 10^{-4}$).

\subsection{Bispectrum}

We now consider the implications of the effective field theory~(\ref{EFT-new-physics-2}) on the bispectrum. The main interactions leading to new effects are due to $M_2^4$ and $M_3^4$ in the action~(\ref{modified-action-2}), and are given by 
\bea
\mathcal L_{M_2}^{(3)} &=&  -M_{\rm Pl}^2  a^3 |\dot H|    (\tilde \nabla \pi)^2   \Sigma (\nabla^2) \dot \pi    ,  \\
\mathcal L_{M_3}^{(3)} &=& M_{\rm Pl}^2  a^3 |\dot H|     \frac{2 M_3^{4}  c_{\rm s}^2 }{3M_2^4 (1 - c_{\rm s}^2 )   }   \, \dot \pi   \Sigma (\nabla^2) \left( \dot \pi \Sigma (\nabla^2)\dot \pi \right)    ,
\eea
where $\Sigma$ was defined in eq.~(\ref{Sigma-def}). To obtain an estimate of the size of non-Gaussianity, we proceed by weighting the strength of the quadratic and cubic operators of the theory, evaluated at Hubble crossing in the range (\ref{nprangeH}). Then, $f_{\rm NL}$ is approximately given by the relation
\be \label{eq-ng-zeta}
\frac{\mathcal L^{(3)}}{\mathcal L^{(2)}} \sim f_{\rm NL} \zeta ,
\ee 
where the length scales are replaced by their values during Hubble crossing and $\zeta=-H\pi$. Due to the fact that the operators present in the theory consist of spatial derivatives which decay rapidly on super-horizon scales, we expect~\cite{Cheung:2007st} the dominant contribution to non-Gaussianity to be that where all momenta are of similar magnitude, i.e. of the equilateral type. Therefore, if Hubble crossing happened within the new physics regime, then $\omega^2 = p^4 / ( M c_{\rm s}^{-1} )^2  \simeq H^2$, implying that $p^2 \simeq M H/c_{\rm s}$. This allows us to consider the following replacements when evaluating the ratio (\ref{eq-ng-zeta}):
\be \label{horizon_cross_replacements_nl}
\nabla^2 / a^2 \to M c_{\rm s}^{-1} H , \qquad  \partial_t \to H , \qquad  \Sigma(\tilde\nabla^2) \to \frac{M c_{\rm s}^{-1}}{H} .
\ee
Let us first estimate the contribution of the quadratic piece:

\be
\mathcal L^{(2)} \,\, = \,\, a^3 M_{\rm Pl}^2 |\dot H| \dot \pi  \Sigma(\tilde\nabla^2)  \dot \pi \,\, \simeq  \,\, a^3 M_{\rm Pl}^2 |\dot H|  H \Lambda_{\rm UV} \pi^2  .
\ee 
The cubic contribution coming from the $M_2^4$ piece is given by 
\be
\mathcal L_{M_2}^{(3)} \,\, = \, \, - a^3 M_{\rm Pl}^2 |\dot H|    \frac{(\partial \pi)^2 }{a^2}  \Sigma(\tilde\nabla^2) \dot \pi \,\, \simeq \, \, a^3 M_{\rm Pl}^2 |\dot H|  \Lambda_{\rm UV}^2  \pi^2 ( -H \pi ) .
\ee 
Finally, the contribution coming from the $M_3^4$ piece is given by
\be
\mathcal L_{M_3}^{(3)} =  a^3 \frac{2 M_3^4 c_{\rm s}^2}{3M_2^4(1-c_{\rm s}^2)} M_{\rm Pl}^2 |\dot H|  \bigg[  \dot \pi  \Sigma(\tilde\nabla^2)  \bigg]^2  \dot \pi    \simeq a^3  \frac{2 M_3^4 c_{\rm s}^2}{3M_2^4(1-c_{\rm s}^2)} M_{\rm Pl}^2 |\dot H|    \Lambda_{\rm UV}^2  \pi^2  (H \pi) .
\ee 
Putting the previous results together, we thus see that the generic prediction is
\be
f_{\rm NL} \sim \frac{\Lambda_{\rm UV}}{H} , \label{f-NL-prediction}
\ee
which implies a large non-Gaussian signature. This result shows explicitly that the scale $\Lambda_{\rm UV}$ enters directly into the computation of low-energy observables such as the level of non-Gaussianity.  For instance, if $\Lambda_{\rm UV} \sim \Lambda_{\rm s.b.}$ then we obtain a sizable estimation of order $f_{\rm NL} \sim 10^{2}$. 

We have seen how the magnitude of $f_{\rm NL}$ can be large in the new physics regime where $M^2 \ll p^2 \ll M^2 c_{\rm s}^{-2}$. In addition, although the dominant contribution to the three-point function will still be of equilateral type, we expect that the change in the dispersion relation sources deviations from the equilateral configuration~\cite{Creminelli:2005hu,Ashoorioon:2011eg,Chialva:2011hc}. In our approach though, there is yet another potential source of novel shapes of the three-point functions. This is the scale dependence of the coefficients in the Lagrangian \eqref{modified-action-2} in the non-linear dispersion relation regime, a claim which is currently under investigation.

\section{Conclusions} \label{section-conclusions}
\setcounter{equation}{0}

Our work emphasizes once more the power of the effective field theory perspective for studying inflation~\cite{Cheung:2007st}. 
As we  have seen, action~(\ref{starting-action}) represents the lowest order expansion, in terms of spacetime derivatives, of the most general EFT of inflation driven by a single degree of freedom. As such, it is limited in that for a suppressed speed of sound $c_{\rm s} \ll 1$ its strong coupling scale $\Lambda_{\rm s.c.}$ is found to be much lower than the high-energy cutoff scale $\Lambda_{\rm UV}$ at which new degrees of freedom start participating in the inflationary dynamics.  This limitation is particularly critical in the regime where we might optimistically hope that the EFT formalism provides non-trivial phenomenology through large non-Gaussian signatures. To address this limitation, in this work we have considered an extension to the standard EFT of inflation motivated by previously proposed UV-completions, implying a scale-dependent self-coupling for the Goldstone boson $\pi$. This extension involves $n$-point interactions of the form
\be
\mathcal L_{\rm EFT}^{(n)} \propto   \left[ (1+ g^{00}) \frac{M^2 }{ M^2 - \tilde\nabla^2} \right]^{n-1} (1+g^{00})  , \label{extension-conclusions}
\ee
where $1 + g^{00} \sim - 2 \dot \pi$ in the decoupling limit, which we derive by integrating out heavy fields coupled to $\pi$.  A specific example for a single heavy field was presented in Section \ref{singleheavyfieldexample}, while the general calculation for arbitrarily many heavy fields is given in Appendix  \ref{several-massive-fields}.

The extended EFT implied by (\ref{extension-conclusions}) allows us to access a regime of so-called new physics, characterized by a modified dispersion relation $\omega(p)$ which is quadratic in momentum $p$ above the energy scale $\Lambda_{\rm new} = M c_{\rm s}$, where $M$ is the mass of the field being integrated out. We have shown that, as emphasized in ref.~\cite{Baumann:2011su}, this modified dispersion relation has important consequences for any physical process sensitive to the scaling properties of the various operators appearing in the low-energy EFT, including the non-linear self-interaction of the $\pi$ field. We thus have an EFT description which is valid throughout the low-energy regime, and which is fully consistent with the new physics regime in such a way that the modified dispersion relation is realized by non-linear interactions at all orders in perturbation theory. This generalizes the analysis of \cite{Baumann:2011su} by  demonstrating that the new physics regime may be fully incorporated within the EFT formalism from the very beginning, without the need for an {\it ad hoc} completion to keep the theory weakly coupled. In this particular respect, we have shown explicitly that the scaling properties induced by (\ref{extension-conclusions}) 
raise the strong coupling scale above the UV cutoff scale $\Lambda_{\rm UV}$ (Section \ref{thestrongcouplingscale}). This is in part due to the fact that the extended effective field theory~(\ref{modified-action-2}) captures accurately the non-trivial role that UV physics has on the Goldstone boson as the energy increases, without implying the existence of extra degrees of freedom. In fact the single field EFT description is sensible exactly as far as expected, that is until the heavy degrees of freedom of the UV complete theory are excited, as we showed in Section \ref{UVdof}.

To summarize our results concerning the phenomenology of inflation, we find that the extended effective field theory~(\ref{modified-action-2}) predicts the following relations between cosmological observables and parameters:
\be
\mathcal P_{\zeta}  \simeq \frac{2.7}{100} \frac{H^2}{M_{\rm Pl}^2 \epsilon} \sqrt{\frac{\Lambda_{\rm UV}}{H}} , \qquad   r \simeq 7.6 \epsilon  \sqrt{\frac{H}{\Lambda_{\rm UV}} } ,  \qquad f_{\rm NL} \sim \frac{\Lambda_{\rm UV}}{H} , \label{predictions-1}
\ee
where $\Lambda_{\rm UV} = M c_{\rm s}^{-1}$. Although the speed of sound $c_{\rm s}$ does not appear explicitly in these expressions, a suppressed value of it is crucial for the new physics regime to exist. These predictions may be compared to the ones obtained in the case of the standard effective field theory~(\ref{starting-action}), given by
\be
\mathcal P_{\zeta}  \simeq \frac{1.3}{100} \frac{H^2}{M_{\rm Pl}^2 \epsilon c_{\rm s}}  , \qquad   r \simeq 16 \epsilon c_{\rm s} ,  \qquad f_{\rm NL} \sim \frac{1}{c_{\rm s}^2} ,  \label{predictions-2}
\ee
which are also compatible with the effective field theory~(\ref{modified-action-2}) in the regime $\omega^2 \ll \Lambda_{\rm new}^2 (=\Lambda_{\rm UV}^2 c_{\rm s}^{4})$. Apart from minor numerical factors, the two sets of predictions differ in their dependence on the theoretical parameters characterizing the two regimes, such that one may replace
\be
c_{\rm s}^2 \to \frac{H}{\Lambda_{\rm UV}} ,
\ee
to go from (\ref{predictions-2}) to  (\ref{predictions-1}). These results suggest that one should be careful when interpreting future results from surveys relevant for constraining inflation.\footnote{Recall that other possible parametrizations for these observables, arising from different operators in the Lagrangian, are given in \cite{Cheung:2007st, Senatore:2009gt, Senatore:2010jy}, for example.}  Adopting an optimistic perspective, an observation of large values for $r$ and $f_{\rm NL}$ in the near future would allow us to infer the values of the parameters $\{\epsilon,H,c_{\rm s}\}$ in the case we assume the validity of theory~(\ref{starting-action}) and  $\{\epsilon,H,\Lambda_{\rm UV}\}$ in the case of theory~(\ref{modified-action-2}). If $f_{\rm NL}$ turns out to be small, it would be impossible to infer the existence of a new physics regime, and we would be forced to consider the theory~(\ref{starting-action}) as the best parametrization of inflation. However, a large value of  $f_{\rm NL}$ would open up the possibility that Hubble crossing happened within the new physics regime, implying a drastically different interpretation of the available data. In such a case, it would be preferable to consider a parametrization of inflation consistent with a weakly coupled description, like the one offered by the extended version~(\ref{modified-action-2}).

It is clear that the examples considered in this work constitute a subset of new physics modifications compatible with a consistent EFT formulation. While we expect that any new physics extension would share similar characteristics to those discussed here, it could be important to re-examine other existing representations motivated by different UV completions, such as DBI-inflation~\cite{Alishahiha:2004eh}. Even in the present case, there are observational consequences we have not considered here at all.  In terms of the three-point function, many detailed questions now arise:
\begin{itemize}
\item While we have assumed equilateral type non-Gaussianity and calculated only approximately the magnitude of $f_{\rm NL}^{\rm (eq)}$, it is possible that a more careful calculation will reveal deviations from the equilateral case arising from the modified dispersion relation or sourced by the modified scale-dependent interactions in the theory. 

\item In estimating $f_{\rm NL}^{\rm (eq)}$, we also assumed that there were modes which crossed the Hubble horizon in the range (\ref{nprangeH}), and that $M_2$ and $M_3$ were nonzero. It is a non-trivial question whether there exist explicit examples, with physically motivated parameters $M_n$, for which this regime will give rise to observable non-Gaussianities. 

\item It would be interesting to analyze how the present effective field theory could affect other types of non-Gaussianity, so far analyzed in the context of the parametrization~(\ref{starting-action}), including the so called local-~\cite{Senatore:2010wk} and resonant-type~\cite{Behbahani:2011it, Behbahani:2012be}. 
\end{itemize}
Last but not least, the modifications studied here could even be relevant to study new physics for tensor modes. The effect of the extrinsic curvature terms in~(\ref{starting-action}) is limited to a dispersion relation of the form $\omega (p) = c_{\rm h} \,  p$ (where $c_{\rm h} = 1 - \bar M_3^2 / M_{\rm Pl}^2$ denotes the speed of sound for tensor-modes), in contrast to the modified dispersion relation they imply for scalars. However, for instance, if tensor modes interacted with additional spin-2 fields during inflation, their dispersion relation would be non-trivially modified~\cite{Atal:2011te}, implying an extended EFT with extra spacetime differential operators affecting the extrinsic curvature terms of~(\ref{starting-action}).

\subsection*{Acknowledgements}

We would like to thank Ana Ach\'ucarro, Neil Barnaby, Daniel Baumann, Sebasti\'an C\'espedes, Fotis Diakonos, Jinn-Ouk Gong, Subodh Patil and the participants of the ``Effective Field Theory in Inflation'' workshop (at the Lorentz Center, Leiden) for useful discussions and comments on this work.  
The work of GAP was supported by funds from Conicyt under both a Fondecyt ``Iniciaci\'on'' project 11090279 and an ``Anillo'' project ACT1122. 
GAP wishes to thank King's College London, and the University of Cambridge (DAMTP) for their hospitality during the preparation of this work. RG was supported by an SFB fellowship within the Collaborative Research Center 676 ``Particles, Strings and the Early Universe''
and would like to  thank the theory groups at DESY and the University of Hamburg for their
 hospitality while this work was being carried out. RG is also grateful for the support of the European Research Council via the Starting Grant numbered 256994.
\begin{appendix}

\renewcommand{\theequation}{\Alph{section}.\arabic{equation}}
\setcounter{section}{0}
\setcounter{equation}{0}

%

\section{Integration of several massive fields} \label{several-massive-fields}
\setcounter{equation}{0}
Here we justify the form of the interaction terms that appear in the generalized effective action \eqref{basic-modification}, upon integrating out several heavy fields.
Let us write the simplest action coupling multiple heavy fields to $\delta g^{00} \equiv g^{00} + 1 $. We are interested in extracting tree-level effects, and therefore we consider an action quadratic in the heavy fields, but to all orders in $\delta g^{00}$. To lowest order in $\delta g^{00}$, we have
\be 
S = -\frac{1}{2}\int d^3xdt \sum_a  \Bigg\{     \mathcal F_a \left[  -\Box \mathcal   + M_a^2 -  B_a (g^{00} + 1)  \right]  \mathcal F_a +  2 A_a (g^{00} + 1) \mathcal F_a   + \sum_{b} C_{ab}  (\mathcal F_a \dot{ \mathcal F_b} )  \Bigg\} ,
\ee 
where $A_a$, $B_b$ and $C_{ab}$ are background quantities. The matrix $C_{ab}$ is an anti-symmetric matrix, $\Box$ corresponds to the FLRW version of the D'Alambertian operator 
\be
\Box =  -\partial_t^2 - 3H\partial_t + \tilde\nabla^2 ,
\ee
and the couplings have mass dimensions $[A]=3, \ [B]=2, \ [C]=1$. Notice that we have excluded non-diagonal mass terms, which may be eliminated by field redefinitions.

To proceed, we neglect the friction terms coming from the volume factor $a^3$ in $d^3xdt$, and focus on the general structure stemming from integrating out the massive fields $\mathcal F_a$. The more elaborate case in which the friction term is incorporated is completely analogous. The equations of motion are:
\be 
 (  - \Box + M_a^2 - B_a  (g^{00} + 1) ){\mathcal F_a} + \sum_b C_{a b} \dot {\mathcal F_b} = -A_a (g^{00} + 1) .
\ee
We are interested in the low-energy behavior of this system. Therefore, following the reasoning of Section~\ref{integration-1-massive-field}, we disregard the time derivative $\partial_t^2 + 3 H \partial_t$ when compared to the operator $M_a^2 - \nabla^2$. On the other hand, we do not neglect the time derivative in the interaction term, as its role is to couple different massive fields, and its contribution depends on the strength of $C_{ab}$. These considerations lead to the equation
\be 
\Omega_a {\mathcal F_a} + \sum_b C_{a b} \dot {\mathcal F_b} = -A_a (g^{00} + 1) , \label{Lagrange-heavy-fields}
\ee
where
\be \label{omegaexpansion}
\Omega_a \equiv  M_a^2 - \nabla^2  - B_a  (g^{00} + 1) .
\ee
Since in this limit the heavy fields $\mathcal F_a$ are non-dynamical, we may treat them as Lagrange multipliers and insert them back into the action without kinetic terms. This leads to the EFT action contribution due to the heavy fields:
\be 
S = -\frac{1}{2}\int d^3xdt \sum_a     (g^{00} + 1)  A_a \mathcal F_a ,
\ee 
where the $\mathcal F_a  $ are the solutions of (\ref{Lagrange-heavy-fields}) which we now proceed to obtain. First, notice that (\ref{Lagrange-heavy-fields}) may be re-expressed as:
\be
\left(\begin{array}{cccc} \Omega_1 & C_{12} \partial_t & C_{13} \partial_t & \cdots \\ - C_{12} \partial_t & \Omega_2 & C_{23} \partial_t & \cdots \\ - C_{13} \partial_t & - C_{23} \partial_t &  \Omega_3   & \cdots \\ \vdots &  \vdots & \vdots &  \ddots  \end{array}\right) \left(\begin{array}{c} \mathcal F_1 \\ \mathcal F_2 \\ \mathcal F_3 \\  \vdots \end{array}\right) = -\left(\begin{array}{c} A_1 \\ A_2 \\ A_3 \\  \vdots \end{array}\right) (g^{00} + 1) .
\ee
To deal with this equation, we assume that the off-diagonal terms are subleading when compared to the diagonal terms $\Omega_a$. This allows us to invert the matrix operator perturbatively, leading to the first-order result:
\be \label{heavy-eom}
 \left(\begin{array}{c} \mathcal F_1 \\ \mathcal F_2 \\ \mathcal F_3 \\  \vdots \end{array}\right) = -
 \left(\begin{array}{cccc} \Omega_1^{-1} & - \Omega_1^{-1} C_{12}  \partial_t  \Omega_2^{-1} & - \Omega_1^{-1} C_{13} \partial_t \Omega_3^{-1} & \cdots \\  \Omega_2^{-1} C_{12} \partial_t  \Omega_1^{-1} &  \Omega_2^{-1} & - \Omega_2^{-1} C_{23} \partial_t \Omega_3^{-1}  & \cdots \\  \Omega_3^{-1} C_{13} \partial_t \Omega_1^{-1} &  \Omega_3^{-1} C_{23} \partial_t  \Omega_2^{-1} & \Omega_3^{-1}   & \cdots \\ \vdots &  \vdots & \vdots &  \ddots  \end{array}\right)
 \left(\begin{array}{c} A_1 \\ A_2 \\ A_3 \\  \vdots \end{array}\right)  (g^{00} + 1)  ,
\ee
which may be re-expressed as:
\be
\mathcal F_a = - \frac{A_a }{\Omega_a } (g^{00} + 1) + \sum_{b} C_{a b}  \frac{1}{\Omega_a} \partial_t   \frac{1}{\Omega_b} A_b (g^{00} + 1) .
\ee
We may now plug this solution back into the action, obtaining:
\be 
S = \frac{1}{2}\int d^3xdt   \left\{   \sum_a A_a  A_a   (g^{00} + 1) \frac{ 1}{\Omega_a } (g^{00} + 1) - \sum_{ab}A^a C_{a b}  (g^{00} + 1)  \frac{1}{\Omega_a} \partial_t   \frac{1}{\Omega_b} A^b (g^{00} + 1)  \right\} .
\ee 
To simplify this expression notice that, due to the anti-symmetry of $C_{ab}$, the second term vanishes whenever the time derivative $\partial_t$ acts on a quantity that does not carry the label $b$. This means that the only non-vanishing contributions coming from the second term are those proportional to $\dot A_b$, $\dot B_b$ and $\dot M^2_b$. For definiteness, and to keep our discussion simple, let us assume that both $B_b$ and $M_b^2$ are constants and consider only a time dependence of the $A_a$ coefficients. In this case, we obtain the formal result:
\be 
S = \frac{1}{2}\int d^3xdt   \left\{  \sum_a A_a  A_a (g^{00} + 1)  \frac{ 1 }{\Omega_a } (g^{00} + 1) - \sum_{a b} ( C_{a b} A_a \dot A_b ) (g^{00} + 1) \frac{1}{\Omega_a  \Omega_b}   (g^{00} + 1)  \right\} .
\label{action-for-integrated-heavy-fields}
\ee 
As discussed in Section~\ref{integration-1-massive-field}, the inverse of $\Omega_a$ is an operator which has the following expansion:
\bea
\Omega_a^{-1} &=&  \frac{1 }{M_a^2 -  \nabla^2} \left[1  -  (g^{00} + 1) \frac{ B_a }{M_a^2 -  \nabla^2} \right]^{-1} \nn \\
&=&   \frac{1 }{M_a^2 -  \nabla^2} \sum_n \left[   (g^{00} + 1) \frac{ B_a }{M_a^2 -  \nabla^2} \right]^{n}.
\eea
Then, inserting this expansion back into the action (\ref{action-for-integrated-heavy-fields}) and keeping  terms up to cubic order, we finally arrive at the expression
\bea
S &=& \frac{1}{2}\int d^3x dt   \Bigg\{  (g^{00} + 1) \left[  \sum_a  \frac{ A_a^2 }{M_a^2 - \nabla^2 }  - \sum_{a b}  \frac{C_{a b}A_a \dot A_b }{( M_a^2 - \nabla^2 )( M_b^2 - \nabla^2 )} \right] (g^{00} + 1)  \nn \\  
&& 
+ \sum_a  A_a^2 B_a  (g^{00} + 1)  \frac{ 1 } {M_a^2 - \nabla^2 }   \left[ (g^{00} + 1)  \frac{ 1 } {M_a^2 - \nabla^2 }  (g^{00} + 1) \right]   \nn \\ 
&& - \sum_{a b } C_{a b}A_a \dot A_b B_b (g^{00} + 1)  \frac{1}{M_a^2 - \nabla^2 }  \left[ (g^{00} + 1) \frac{1}{( M_b^2 - \nabla^2 )( M_c^2 - \nabla^2 )}   (g^{00} + 1)  \right] \nn \\ 
&& - \sum_{a b } C_{a b}A_a \dot A_b B_a  (g^{00} + 1) \frac{1}{( M_b^2 - \nabla^2 )( M_c^2 - \nabla^2 )}   \left[   (g^{00} + 1)  \frac{1}{M_a^2 - \tilde \nabla^2 }  (g^{00} + 1)  \right]     
 + \cdots  \Bigg\} . \nn\\
\label{action-for-integrated-heavy-fields-2}
\eea
This implies that the general quadratic action for the Goldstone boson $\pi$ takes the form
\bea
\nonumber S^{(2)} & = & - M_{\rm Pl}^2 \int d^3 x dt a^3 \dot H \bigg[\dot \pi \bigg (1 +  \sum_a \frac{\beta_a}{ M_a^2 - \tilde \nabla^2}  +  \sum_{ab}\frac{ \beta_{ab} }{( M_a^2 - \tilde \nabla^2 )(  M_b^2  - \tilde \nabla^2) }   + \cdots \bigg )\dot \pi \\&&  - ( \tilde \nabla \pi)^2  \bigg],
\eea
where $\beta_a$ parametrizes the coupling to a heavy field with index $a$, and $\beta_{ab}$ parametrize the interactions between heavy fields carrying labels $a$ and $b$ etc.
In momentum space the action takes the form
\bea
\nonumber S^{(2)} & = & - M_{\rm Pl}^2 \int d^3 k dt a^3 \dot H \bigg[\dot \pi \bigg (1 +  \sum_a \frac{\beta_a}{ M_a^2 + p^2}  +  \sum_{ab}\frac{ \beta_{ab} }{( M_a^2 + p^2)(  M_b^2  + p^2) }   + \cdots \bigg )\dot \pi \\&&  + p^2 \pi^2 \bigg].
\eea
The equation of motion for the $\pi$ field is therefore given by
\be
\ddot \pi + 3 H \dot  \pi - c_{\rm s}^2 (p^2) p^2 \pi = 0 ,
\ee
where
\bea
c_{\rm s}^2(p) & = & {\prod_a (M_a^2 + p^2)} \times \nonumber \\ &&  \left [\!\prod_a (M_a^2 + p^2) \! + \! \sum _a \beta_a \prod_{b \neq a} (M_b^2 \!+\! p^2) \! + \! \sum_{a < b} \beta_{ab} \prod_{c \neq a, b}  (M_c^2 \! + \! p^2) \! + \! \ldots \! + \! \beta_{12\ldots N} \right ]^{-1}\!\!\!\!\!\!\!\!.
\eea
The inverse speed of sound squared is defined as the limit
\be \label{cs_in_multi_int_fields}
c_{\rm s}^{-2}\equiv\lim_{p\rightarrow 0}c_{\rm s}^{-2}(p)=1+\sum _a\frac{ \beta_a}{M_a^2} + \sum_{a < b} \frac{\beta_{ab}}{M_a^2M_b^2} + \ldots + \frac{\beta_{12\ldots N}}{M_1^2 M_2^2 \ldots M_N^2},
\ee
where $N$ is the number of heavy fields and the indices run from $1\ldots N$.
To analyze this, let us consider the short wavelength regime where the friction term can be disregarded and $p\equiv k /a$ may be taken as a constant.  The dispersion relation is then 
\be
\omega^2 (p) = c_{\rm s}^2(p) p^2.
\ee
For the case of one additional heavy field we get
\be
c_{\rm s}^2(p) = (M^2 + p^2)\left[M^2 + p^2 + \beta \right]^{-1},
\ee
which reduces to the expressions in Eq.~\eqref{full-modified-dispersion},~\eqref{full-modified-cs} when $\beta = \dfrac{2M_2^4 M^2}{M_{\rm Pl}^2|\dot H|}$. 
For multiple non-interacting fields where $\beta_{ab\ldots}=0$, this becomes
\be
c_{\rm s}^2(p) = \prod_{a} (M_a^2 + p^2)\left[\prod_{a} (M_a^2 + p^2) + \sum _{a} \beta_a\prod_{b \neq a} (M_b^2 + p^2) \right]^{-1},
\ee
with the inverse speed of sound squared given by
\be \label{cs_in_multi_non_int_fields}
c_{\rm s}^{-2}=1+\sum _a\frac{ \beta_a}{M_a^2}.
\ee
Recall that we are restricted to the low-energy regime $$ \omega^2 \ll M_a^2 + p^2 $$ in order for the expansion (\ref{omegaexpansion}) to be valid. Without loss of generality we can consider two cases: one where the $M_a$ are all comparable, and the other where there exists some hierarchy among these heavy masses. This can be studied using a representative lowest mass $M_l^2$; either other masses are comparable, or significantly larger. In the former case we require the inequality to hold for all $a$, while in the latter just that  $ \omega^2 \ll M_l^2 + p^2 $. 
The generic UV scale for arbitrary number of fields with different masses will be a complicated function of the speed of sound and the mass scales of the problem, so let us study in some detail the case where all the heavy masses $M_a$ are comparable: $M_a^2 \approx M^2 \, \, \, \forall\, \,  a$. The dispersion relation then reads
\bea\label{dispersion_multi_same_mass}
\omega^2(p) & = & (M^2 + p^2) p^2 \times \nonumber \\ &&  \left [ M^2 + p^2 + \sum _a \beta_a + (M^2 + p^2)^{-1}\sum_{a < b} \beta_{ab}  + \ldots + \beta_{12\ldots N}(M^2 + p^2)^{1-N} \right ]^{-1}\!\!\!\!\!.
\eea
From this expression we can read off the low-energy regime as an upper bound in the momentum
\be
\label{LER}
p^2 \ll M^2 + \sum _a \beta_a + (M^2 + p^2)^{-1}\sum_{a < b} \beta_{ab}  + \ldots + \beta_{12\ldots N}(M^2 + p^2)^{1-N}.
\ee
We see that in general this is a polynomial inequality of degree $N$ in squared momentum,
$$G^N(p^2)\ll 0 \ .$$ Therefore the solution is $p^2\ll p^2_{\rm UV}(M,c_{\rm s},\beta)$ with $p^2_{\rm UV}$ representing the degenerate positive root of the polynomial $G^N$.
The energy scale $\Lambda_{\rm UV}$ is then given by substituting $p^2_{\rm UV}$ into the dispersion relation. Since this is the root of the polynomial $G^N(p^2)$ the denominator of Eq.~\eqref{dispersion_multi_same_mass} is just proportional to $p^2_{\rm UV}$ and the expression simplifies to
\be\label{luv_multi}
\Lambda_{\rm UV}^2 \sim (M^2 + p^2_{\rm UV}) \ .
\ee

We also see a modification of the dispersion relation in the multiple heavy field case. For small values of $p^2$ compared to the mass squared, the low-energy regime condition (\ref{LER}) becomes 
\bea
p^2 & \ll &  M^2 +  \sum_a \beta_a + \sum_{a\neq b} \beta_{ab} M^{-2} + .... + \beta_{12....n} M^{2(1-n)}. 
\eea
This inequality is automatically satisfied when $p^2 \ll M^2$. The dispersion relation in this regime becomes
\bea
\label{MFdisprelnlowmom} \omega^2(p) & = & c_{\rm s}^2 p^2  (1 + \frac{p^2}{M^2})^n\, \,,
\eea
where $c_{\rm s}^2$ is given Eq.~\eqref{cs_in_multi_int_fields}. Note that the sound speed is lowered, and lowered additively, by the presence of heavy fields in this regime.  The expansion (\ref{MFdisprelnlowmom}) includes terms dependent on $p^4, p^6...$, but these are suppressed by increasing powers of $p^2/M^2$, so that we recover the usual $\omega^2 \sim p^2$ dispersion relation in this regime. 
For large values of $p^2$ compared to $M^2$, the low-energy condition becomes
\bea
\label{LEClargep} p^{2n} & \ll &   \left ( \sum_a \beta _a p^{2(n-1)} + \sum_{a \neq b} \beta_{ab} p^{2(n-2)} + .... + \beta_{123...n}\right ).
\eea
The dispersion relation in this regime is given by 
\bea
\omega^2(p)& = &  ( \sum \beta_a p^{-4} + \sum \beta_{ab} p^{-6} + ... + \beta_{1...n} p^{-2n - 2})^{-1}.
\eea
We see that many powers of $p$ can enter. However, for large $p^2 \gg M^2$, the subleading terms in $\Lambda(p)$ in the denominator are suppressed, and the dominant $p$-dependence of the dispersion relation is given by 
\be
\omega^2(p) \approx \frac{p^4}{\sum \beta_a}.
\ee

\end{appendix}

\end{document}